\documentclass[prl,twocolumn,notitlepage,superscriptaddress,floatfix]{revtex4-2}

\usepackage{hyperref}
\usepackage{mathtools}
\usepackage{amsmath}
\usepackage{amssymb}
\usepackage{graphicx}
\usepackage[dvipsnames]{xcolor}
\usepackage{braket}
\usepackage[normalem]{ulem}
\usepackage[capitalise]{cleveref}
\usepackage[export]{adjustbox}
\usepackage{makerobust}
\newcommand{\bvec}[1]{\boldsymbol{#1}}
\newcommand{\stkout}[1]{%
    \ifmmode\text{\sout{\ensuremath{#1}}}%
    \else\sout{#1}%
    \fi%
}

\newcommand{\approptoinn}[2]{\mathrel{\vcenter{
  \offinterlineskip\halign{\hfil$##$\cr
    #1\sim\cr\noalign{\kern-1pt}#1\propto\cr\noalign{\kern2pt}}}}}
\newcommand{\appropto}{\mathpalette\approptoinn\relax}

\newcommand{\captiontitle}[1]{{\bf #1}}
\newcommand{\captionlabel}[1]{{\bf(#1)}}
\newcommand{\unsim}{{\sim}}

\newcommand{\makeauthor}[2]{\newcommand{#1}[1]{{%
  \protect%
  \color{#2} ##1}}%
  \MakeRobustCommand#1}

\newcommand{\delete}[1]{{\color{blue}\stkout{#1}}}
\newcommand{\append}[1]{{\color{blue}#1}}

\renewcommand{\delete}[1]{}
\renewcommand{\append}[1]{#1}

\makeauthor{\lk}{orange}
\makeauthor{\af}{blue}
\makeauthor{\dmk}{red}
\makeauthor{\ms}{ForestGreen}
\makeauthor{\lc}{purple}

\newcommand{\WSe}{WSe\textsubscript{2}}
\newcommand{\LambdaC}{\Lambda_\mathrm{c}}

\usepackage{tikz}
\usepackage{forloop}
\makeatletter
\def\parsecomma#1,#2\endparsecomma{\def\page@x{#1}\def\page@y{#2}}
\tikzdeclarecoordinatesystem{page}{
    \parsecomma#1\endparsecomma
    \pgfpointanchor{current page}{north east}
    \pgf@xc=\pgf@x%
    \pgf@yc=\pgf@y%
    \pgfpointanchor{current page}{south west}
    \pgf@xb=\pgf@x%
    \pgf@yb=\pgf@y%
    \pgfmathparse{(\pgf@xc-\pgf@xb)/2.*\page@x+(\pgf@xc+\pgf@xb)/2.}
    \expandafter\pgf@x\expandafter=\pgfmathresult pt
    \pgfmathparse{(\pgf@yc-\pgf@yb)/2.*\page@y+(\pgf@yc+\pgf@yb)/2.}
    \expandafter\pgf@y\expandafter=\pgfmathresult pt
}
\makeatother
\newenvironment{supplement}{\clearpage\pagenumbering{gobble}}{}
\newcommand{\supppage}[2]{%
  \begin{tikzpicture}[remember picture,overlay,inner sep=0pt,outer sep=0pt]
    \node[anchor=north west] at (page cs:-1,1) {\includegraphics[page=#1]{#2}};
  \end{tikzpicture}
  \clearpage
}
\newcommand{\arxivSubmit}[2]{%
\begin{supplement}
  \newcounter{suppctr} \forloop[1]{suppctr}{0}{\value{suppctr} < #2}{%
    \supppage{\the\numexpr\value{suppctr}+1\relax}{#1}%
  }
\end{supplement}
}

\begin{document}
\title{Competition of Density Waves and Superconductivity in Twisted Tungsten Diselenide}

\author{Lennart Klebl}
\author{Ammon Fischer}
\affiliation{Institut f\"ur Theorie der Statistischen Physik, RWTH Aachen University and JARA-Fundamentals of Future Information Technology, D-52056 Aachen, Germany}
\author{Laura Classen}
\affiliation{Max Planck Institute for Solid State Research, D-70569 Stuttgart, Germany}
\author{Michael M. Scherer}
\affiliation{Institut f\"ur Theoretische Physik III, Ruhr-Universit\"at Bochum, D-44801 Bochum, Germany}
\author{Dante M. Kennes}
\affiliation{Institut f\"ur Theorie der Statistischen Physik, RWTH Aachen University and JARA-Fundamentals of Future Information Technology, D-52056 Aachen, Germany}
\affiliation{Max Planck Institute for the Structure and Dynamics of Matter, Center for Free Electron Laser Science, D-22761 Hamburg, Germany}
\date{\today}

\begin{abstract}
Evidence for correlated insulating and superconducting phases around regions of
  high density of states was reported in the strongly spin-orbit coupled van-der
  Waals material twisted tungsten diselenide (t\WSe{}).  We investigate their
  origin and interplay by using a functional renormalization group approach that
  allows to describe superconducting and spin/charge instabilities in an
  unbiased way. We map out the phase diagram as function of filling and
  perpendicular electric field, and find that the moir\'e Hubbard model for
  t\WSe{} features mixed-parity superconducting order parameters with $s/f$-wave
  and topological $d/p$-wave symmetry next to (incommensurate) density wave
  states. Our work systematically characterizes competing interaction-driven
  phases in t\WSe{}  beyond mean-field approximations and provides guidance for
  experimental measurements by outlining the fingerprint of correlated states in
  interacting susceptibilities.
\end{abstract}
\maketitle
\paragraph{Introduction. ---}
The unique control over band structure and interaction parameters in layered van
der Waals material stacks with long-range moir\'e potentials provides an ideal
platform to simulate many-body phenomena, and thus holds the promise to advance
our understanding of correlated states of matter~\cite{moiresim}. Indeed, a
plethora of correlated phases were reported in different moir\'e materials,
ranging, e.g., from superconductivity and correlated insulators in twisted
multi-layer
graphene~\cite{cao2018unconventional,cao2018mott,lu2019superconductors,Cao2020strange,Polshyn2019,yankowitz2019tuning,liu2021tuning,stepanov2020untying,arora2020superconductivity,Zondiner2020,Wong2020,Xie2019spectrosopic,Kerelsky2019maximized,Jiang2019charge,Choi2019correlations,cao2020nematicity,burg2019correlated,Park2021,cao2021large,hao2021electric,kim2021spectroscopic,liu2020tunable,shen2020correlated,cao2019electric,tutuc2019,rubioverdu2020universal,Chen2019ABC,chen2019evidence,chen2020tunable,park2021multi,zhang2021multi,burg2021multi,kerelsky2021moireless,liu2022isospin}
and transition metal dichalcogenides
(TMDs)~\cite{wang2020correlated,ghiotto2021quantum,tang2020,jin2021stripe} over
excitonic
physics~\cite{jin2019observation,2019Natur.574...76W,shimazaki2020strongly} and
generalized Wigner crystals~\cite{regan2020mott} to quantum anomalous Hall
states~\cite{li2021quantum}.

An example that stands out for its control over a large parameter space is the
twisted homo-bilayer TMD Tungsten Diselenide (t\WSe{}), where a correlated
insulator occurs for a broad range of twist angles
($\theta\approx4^\circ\dots5.1^\circ$) as function of carrier density and
interlayer displacement field~\cite{wang2020correlated,ghiotto2021quantum}.
Theoretically, the twist angle and displacement field affect the relative
interaction and kinetic energy scales, but also the location and strength of
singularities in the density of states (van Hove singularities), and it was
pointed out that there is a correspondence between regions of large density of
states and insulating behavior. The additional observation of zero resistance
states~\cite{wang2020correlated} in its immediate vicinity stimulated a debate
about possible superconductivity and the underlying
mechanisms~\cite{PhysRevB.104.195134,wu2022pair}. 

An unbiased investigation of the electronic phases of t\WSe{} has so far
remained elusive. In this letter, we provide such an analysis of the spin-orbit
coupled triangular moir\'e Hubbard model for t\WSe{} in the \delete{weak to} intermediate
coupling regime \append{--- relevant for experimentally accessible twist angle regimes~\cite{andy2021hartree} ---} using functional renormalization group (FRG) techniques. Within
the FRG, all electronic instabilities are treated on equal footing, providing us
with a tool that can resolve the competition of various electronic correlations.
In particular, the FRG can reveal unconventional mechanisms for
superconductivity from repulsive interactions in an unbiased manner for the full \textit{ab-initio} inspired and material-specific t\WSe{} model.
Thereby it substantially goes beyond previous Hartree-Fock studies~\cite{andy2021hartree} and parquet renormalization group approaches~\cite{PhysRevB.104.195134}.

We perform large-scale simulations of the doping and displacement-field
parameter space and find instabilities towards a variety of density waves around
fillings that correspond to Van Hove singularities, which are flanked by pairing
instabilities. The wave vectors of the density waves are generally
incommensurate and evolve with the displacement field as they follow the nesting
vectors of the Fermi surface, which is in line with a previous Hartree-Fock
study concentrating on commensurate cases~\cite{andy2021hartree}. We find that
fluctuations of the density waves mediate attraction in pairing channels of
mixed parity in wide parameter regimes and predict the corresponding SC order to
be either of mixed $s/f$-wave character for strong doping or of mixed $d/p$-wave
character for moderate doping with a preference to form topological $d+id/p+ip$
combinations in the ground state.

\paragraph{Model. ---} 
The moir\'e band structure of twisted bilayer WSe${}_2$ in a finite out-of-plane
electrical field features a pair of narrow, isolated, and spin-split bands close
to the Fermi level. They are formed by states near valley $K$ or $K'$ of the top
and bottom layer of \WSe{}, which possess opposite spin orientation due to
strong spin-orbit coupling and effective spin-valley locking. As a result,
$SU(2)$ spin symmetry is broken, and the moir\'e band structure reacts strongly
to the potential difference between the layers from a displacement field. This
can be effectively captured by the moir\'e Hubbard
model~\cite{PhysRevLett.122.086402,PhysRevResearch.2.033087,andy2021hartree,doi:10.1063/5.0077901}
\begin{align}\label{eq:Hamiltonian}
    H = -2t\!\!\sum_{\bvec k, m,\sigma}\!\! \cos\!\big(\bvec
    k\!\cdot\!\bvec a_m\!+\! \sigma\varphi\big)\,c^\dagger_{\bvec k,\sigma}
    c^{\phantom{\dagger}}_{\bvec k,\sigma}\!+  U\!\sum_{i} n_{i,\uparrow}
    n_{i,\downarrow}\,,
\end{align}
on the triangular moir\'e lattice with $120^\circ$ nearest-neighbor vectors
$\bvec a_{m=1,2,3}$, describing moir\'e-band electrons
$c^{(\dagger)}_{\bvec{k},\sigma}$ with wave-vectors $\bvec k$ and spin
projection $\sigma \in \{\uparrow, \downarrow\}$. Due to spin-valley locking, $\sigma$ not only describes the spin, but also the valley degree of freedom. The effect of the displacement
field is modeled via a spin-dependent nearest-neighbor hopping $t\,e^{i\sigma
\varphi}$ with absolute value $t$ and phase $\varphi$. Note that the inversion symmetry from the moir\'e lattice leads to an emergent spin-rotational symmetry at zero displacement field $\varphi=0$, despite the strong spin-orbit coupling of the individual \WSe{} layers~\cite{andy2021hartree}.
%
%Moreover, \emph{ab-initio} data show that $\varphi\in[0,\pi/3]$ resembles realistic values of displacement field $D$, {\color{blue}and that $\varphi\appropto D$}~\cite{wang2020correlated,supplement}.
\delete{Moreover, \emph{ab-initio} data show that $\varphi\in[0,\pi/3]$ resembles realistic values of displacement field $D$~\cite{wang2020correlated}.}
\append{Moreover, \emph{ab-initio} data~\cite{wang2020correlated} and atomistic tight-binding simulations (see SM~\cite{supplement}) show that $\varphi\in[0,\pi/3]$ resembles realistic values of displacement field $D$, and that $\varphi\appropto D$.}
The Hubbard interaction
$U$ dominates the Coulomb interaction~\cite{PhysRevResearch.2.033087} and
non-local short-ranged interactions can be screened via substrate
engineering~\cite{Wu17}.

\begin{figure}%
    \centering
    \includegraphics{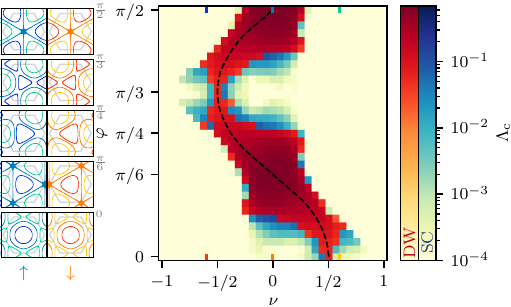}
    \caption{\captiontitle{FRG phase diagram of moir\'e-Hubbard model for
    t\WSe.} We plot the critical scale $\LambdaC$ of the FRG flow that
    corresponds to an onset temperature of the corresponding correlations and
    vary the filling factor $\nu$ and effective displacement field $\varphi$.
    The panels on the left display Fermi surfaces for
    $\varphi\in\{0,\pi/6,\pi/4,\pi/3,\pi/2\}$ (bottom to top), both spin
    polarizations (left: $\sigma=\uparrow$, right: $\sigma=\downarrow$), and
    three values of $\nu\in\{-0.6,0,0.6\}$. The employed FRG approach resolves
    whether the system tends to order in a spin/density wave (DW) or
    superconducting (SC) state, which is encoded as color. Blue regions
    correspond to SC phases with high $\LambdaC$ and red regions correspond to
    DW phases with high $\LambdaC$. Yellow regions show no ordering tendency
    within our approximations and thus are predicted to remain metallic. The
    center of the DW region corresponds to the position of the van-Hove
    singularity for each $\varphi$, indicated by the dashed black line. SC
    phases emerge upon doping slightly away from the DW states.}
    \label{fig:phase-diagram}
\end{figure}

\paragraph{Method. ---}
To study competing phases in this triangular lattice moir\'e Hubbard model, we
employ the functional renormalization group (FRG) and identify the leading
Fermi-surface instabilities including different types of density wave and
superconducting instabilities on equal footing. We use an approximation which
exclusively focuses on the FRG flow of the spin-dependent two-particle
interaction vertex $\Gamma^{(4)}$. Technically, the FRG introduces a scale
parameter $\Lambda$ to interpolate smoothly from the free theory at
$\Lambda=\infty$ to the interacting one at $\Lambda=0$. Ordering tendencies are
indicated by a divergence of $\Gamma^{(4)}$ at finite $\Lambda=\LambdaC$, where,
with our choice of regulator, $\LambdaC$ corresponds to the onset temperature of
strong correlations. Using the effective vertex at the critical scale $\LambdaC$
we can classify the ordering tendencies straightforwardly either as spin/charge
density waves (DW) or as superconductors (SC).
For the present system, we have
extended the standard correlated-electron FRG scheme~\cite{Metzner2012a}:
(1)~the Hamiltonian in Eq.~\eqref{eq:Hamiltonian} does not possess an
$SU(2)$-spin invariance and we have adapted the FRG equations accordingly and
(2)~instead of the widespread scheme of discretizing only wave-vectors on the
Fermi surface, we have employed a scheme in which we finely resolve the full
Brillouin zone (BZ). This facilitates to also resolve incommensurate
density-wave ordering. We note that the latter extension requires a highly
efficient numerical implementation to be able to handle the
$\unsim3.06\times10^9$ coupled ordinary differential equations for the
interaction vertex. For details of the FRG implementation and the analysis of
phases, see~\cite{supplement}.

\paragraph{Phase diagram. ---} \Cref{fig:phase-diagram} summarizes the main results at intermediate interaction strength $U=6t \lesssim 0.7W$ (with the bandwidth $W$) as a function of the filling~$\nu$ and field-dependent phase~$\varphi$. Here $\nu=-1$ corresponds to completely empty, $\nu=0$ to half-filled, and $\nu=1$ to completely filled moir\'e bands.
We adjust the filling by adding a chemical potential term to the Hamiltonian and the given values refer to the filling fraction of the single-particle dispersion.
Upon varying $\varphi$, the DW instabilities follow the location of the Van Hove singularity (VHS). The DW region is most extended around $\varphi=\pi/6$ and $\nu=0$, where the system has a higher-order VHS~\cite{shtyk2017electrons,andy2021hartree}. Given the significant enhancement of density of states at the higher-order VHS as well as the nesting property of the Fermi surface, the DW instability there occurs at high critical scale and in an extended filling region. At the borders of the DW region, superconducting (SC) order emerges. 
The size of the SC regions strongly varies with $\nu$ and $\varphi$. 
Remarkably, for the $\nu=0$  vertical line, i.e. at half filling, we predict SC order for a substantial fraction of values of $\varphi$, interrupted by similarly dominant DW regions. Our findings support the intuitive picture that unconventional SC is driven by the strong spin and charge fluctuations close to the DW instabilities, which we can clearly see in the evolution of the vertex as a function of the RG scale (see SM~\cite{supplement}).

\begin{figure}
    \centering
    \includegraphics{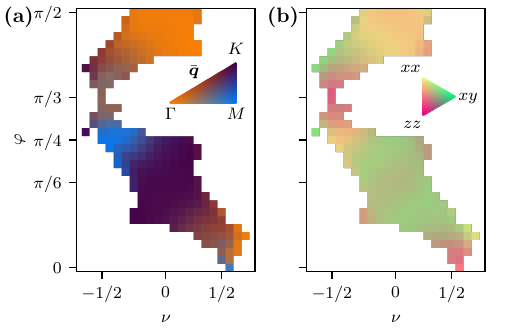}
    \caption{\captiontitle{Momentum and spin structure of spin/density wave
    phases.} \captionlabel{a}~Dominant transfer momentum $\bar{\bvec q}$ of particle-hole
    susceptibility color-coded for all DW instabilities in the phase diagram.
    There are pronounced regions of commensurate ordering vectors: $\Gamma$ at
    $\varphi=\pi/2$ (orange), $K$ at $\varphi=\pi/6$ (dark purple), and $M$ at
    $\varphi=\pi/4$ (blue). The connecting regions in between show
    incommensurate ordering vectors. Note the $SU(2)$ symmetric point
    $\varphi=0$ where $M$ is the dominant ordering vector. \captionlabel{b}~Spin
    structure of particle-hole susceptibility. The relative weight of the
    $\chi^{xx}$ (yellow), $\chi^{xy}$ (green), and $\chi^{zz}$ (pink) components
    is shown for the same DW instabilities as in~\captionlabel{a}. All other
    nonzero spin components of the physical susceptibility are symmetry
    equivalent to either $\chi^{xx}$, $\chi^{xy}$, or $\chi^{zz}$ -- spin rotational symmetry around the $z$-axis implies $\chi^{xx}=\chi^{yy}$ and $\chi^{xy}\propto\chi^{yx}$. For zero
    electric field (i.e. $\varphi=0$), the system is isotropic in spin space
    thus showing perfect degeneracy of the $xx$ ($yy$) and $zz$ components. Upon
    increasing $\varphi$, the $zz$ component is strengthened and then giving way
    to a large region of $xy$-plane ($xx$, $xy$) ordering. At $\varphi=\pi/4$,
    the $xy$ ($yx$) component is weakened and for slightly larger $\varphi$
    giving rise to $zz$ ordering. For $\varphi=\pi/2$, the system favors $xx$
    ($yy$) and $zz$ correlations.} \label{fig:dw-phases}
\end{figure}

\paragraph{Density-wave states. ---}
The strong effect of the displacement field on the band structure also leads to
a changing Fermi surface with varying $\varphi$. In turn, the singular
scattering processes of the DW instabilities correspond to modified wave-vector
transfers. To resolve this evolution in detail, we characterize the momentum and
spin structure of the DW states, see \cref{fig:dw-phases},
and calculate the particle-hole susceptibilities
\begin{align}
  \nonumber
  \chi^D_{\sigma_1 \dots \sigma_4}(\bvec q) &{}= \!\!\!\includegraphics[valign=c]{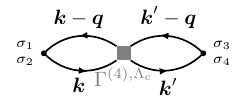} \\
  &{}=\begin{multlined}[t]
      N_{\bvec k}^{-2} \!\!\!\!\!\!
      \sum_{\sigma_{1'}\sigma_{2'}\sigma_{3'}\sigma_{4'},\bvec k,\bvec k'}
      \chi^{0, \LambdaC}_{\sigma_{1^{\vphantom{\prime}}} \sigma_{2^{\vphantom{\prime}}} \sigma_{1'}\sigma_{2'}}(\bvec q,\bvec k) \times{} \\
      %\Gamma^{D,\LambdaC}_{\sigma_{1'}\sigma_{2'}\sigma_{3'}\sigma_{4'}} (\bvec q, \bvec k, \bvec k')
      \hspace{-40pt}\Gamma^{(4),\LambdaC}_{\sigma_{2'},\sigma_{3'},\sigma_{1'},\sigma_{4'}}(\bvec k,\bvec k'-\bvec q,\bvec k-\bvec q)
      \chi^{0, \LambdaC}_{\sigma_{3'}\sigma_{4'}\sigma_{3^{\vphantom{\prime}}}\sigma_{4^{\vphantom{\prime}}}}(\bvec q,\bvec k')\,,
    \end{multlined}
  \label{eq:susc-ph-4pt} \\
  \chi^{ij}(\bvec q) &{}= \sum_{\sigma_1\sigma_2\sigma_3\sigma_4}
  \sigma_i^{\sigma_2\sigma_1}\,
  \chi^D_{\sigma_1\sigma_2\sigma_3\sigma_4}(\bvec q)\,
  \sigma_j^{\sigma_3\sigma_4}
  \label{eq:susc-ph-ij}
\end{align}
for all DW-state regions in \cref{fig:phase-diagram}. Here,
$\chi^D_{\sigma_1,\dots,\sigma_4}$ denotes the four-point particle-hole
susceptibility and $\chi^{ij}$ is its projection to the physical channels
$i,j\in\{0,x,y,z\}$, where $0$ and $x,y,z$ denote charge and spin,
respectively. In Eq.~\eqref{eq:susc-ph-4pt}, we use the
four-point vertex $\Gamma^{(4),\LambdaC}_{\sigma_1\dots\sigma_4}(\bvec k_1,\bvec k_2,\bvec k_3)$ at the critical scale
$\Lambda=\LambdaC$ and contract with the non-interacting particle-hole
susceptibility $\chi^{0, \Lambda}_{\sigma_1 \dots \sigma_4}(\bvec q,\bvec
k)$~\cite{supplement} in order to account for the cross-channel feedback
generated during the FRG flow.

To identify the leading order vector $\bar{\bvec q}$, we sum out the spin
indices of the four-point susceptibility and make a weighted average with the
momentum transfer vector. 
We complement the analysis of
$\bar{\bvec q}$ in \cref{fig:dw-phases}~\captionlabel{a} with a map of the
dominant spin-spin correlations in \cref{fig:dw-phases}~\captionlabel{b}.
By symmetry only three inequivalent spin-spin correlations can be nonzero:
$\chi^{xx}=\chi^{yy}$, $\chi^{xy}=-\chi^{yx}$, and $\chi^{zz}$. Moreover, we find that density-density correlations are subleading across the phase diagram. Nevertheless, the spin-orbit coupling of the system leads to coupled spin- and charge density-waves.
These DW instabilities describe symmetry breaking in the spin- and valley degrees of freedom at the same time owing to spin-valley locking.
For further details on the averaging procedures and density-density correlations, see Ref.~\onlinecite{supplement}.

We find that for the region close to $\varphi=\pi/2$ the system exhibits a
leading ordering vector of $\bar{\bvec q}=\Gamma$, suggesting a ferromagnetic
ground state. The weight is almost equally distributed in $xx/yy$ and $zz$
direction. Moving towards smaller $\varphi$ and following the VHS, the leading
transfer momentum continuously transitions to an extended region around
$\varphi\approx \pi/3$ where $\bar{\bvec q}$ is incommensurate and accompanied
by a strong $zz$ component. Lowering $\varphi$ further to around $\varphi
\lesssim \pi/4$, the support for $xx/yy$ correlations is enhanced and the
dominant ordering vector is $\bar{\bvec q}\sim M$, indicating an instability consistent with the stripe order found in Ref.~\onlinecite{andy2021hartree} or with a more complex superposition of the spin DWs with the three nonequivalent $M$ points as wave vectors~\cite{PhysRevLett.108.227204,PhysRevLett.101.156402}.
Approaching the higher-order VHS at $\varphi=\pi/6$ we see a leading momentum of
$\bar{\bvec q}=K$ and a change towards $xy$ correlations. Notably, for
this choice of $\varphi$, the wave-vector $K$ (as well as $K'$) is a nesting vector connecting the
spin-up with the spin-down Fermi surface. These features signal a twofold
degenerate instability that supports the spiral $120^\circ$-order found in
Ref.~\onlinecite{andy2021hartree}. An analogous signature is visible in the two
small regions at minimal doping at $\varphi\sim\pi/4$ and $\varphi\sim3\pi/8$.
Eventually, letting $\varphi$ go to zero, the ordering vector continuously
approaches $\Gamma$, except for a very small region around $\varphi=0$, i.e.
the limit of restored spin-$SU(2)$ invariance, where $\bar{\bvec q}=M$. The
spin-spin correlations display a slightly more continuous transition towards
$xx/yy$ and $zz$ order at $\varphi=0$, consistent with recovered $SU(2)$
symmetry. 
The feature at $\varphi=0$ is in agreement with previous results for
the spin-$SU(2)$ invariant triangular-lattice Hubbard
model~\cite{PhysRevB.68.104510,scherer2021,gneist2022competing,PhysRevX.11.041013}.

Additionally, we observe that the regions in the phase diagram characterized by a
leading momentum of $\Gamma, M$, or $K$ are connected by extended regions where
the leading momentum is incommensurate.
While the commensurate regions are in
agreement with previous Hartree-Fock studies~\cite{andy2021hartree}, the
unbiased identification of regions with leading incommensurate momentum which can be readily read off from \cref{fig:dw-phases}~\captionlabel{a} is one
of the advantages of our FRG approach featuring high momentum
resolution.

\paragraph{Superconductivity. ---}
In the vicinity of the DW ordered states, our FRG approach can detect pairing instabilities driven by spin and charge fluctuations in an unbiased way because particle-hole and pairing channels are coupled. The corresponding SC states may be classified by the symmetry of the order parameter. We use a linearized gap equation with the vertex at the critical scale $\Lambda=\LambdaC$ (and set the temperature to $T=\LambdaC$) to obtain the pairing gap functions and their respective amplitudes. As for $\varphi\neq0$ the system does not obey $SU(2)$ symmetry, we transform the gap $\Delta_{\sigma\sigma'}(\bvec k)$ to its singlet [$\psi(\bvec k)$] and triplet [$\bvec d(\bvec k)$] components~\cite{sigrist1991phenomenological}. These are inherently coupled giving rise to mixed-parity (singlet and triplet) SC order. Spin rotational symmetry around the $z$ axis mandates that $d_x=d_y=0$ for coupled singlet/triplet instabilities. 
%----------------
\begin{figure}
    \centering
    \includegraphics{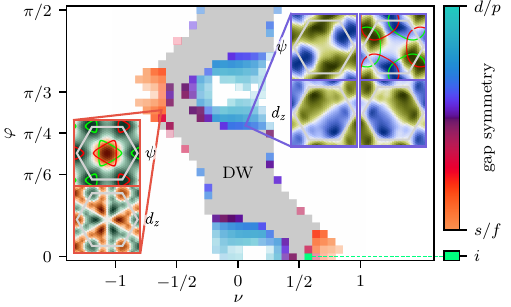}
    \caption{\captiontitle{Properties of the superconducting phases.} The regions of superconducting (SC) order are color-coded by their dominant gap symmetry, with cyan standing for $d/p$-wave SC and orange for (extended) $s/f$-wave SC. Lower values of $\LambdaC$ are indicated by increasing transparency. We plot the logarithmic ratio of $s/f$-wave and $d/p$-wave eigenvalues of the linearized gap equation as a continuous color-bar to highlight regions of strong competition (purple). The small area of bright green denotes $i$-wave SC. In the gray region, FRG predicts spin/density wave order. As for most parts of the phase diagram the system is not $SU(2)$ symmetric, singlet ($\psi$) and triplet ($d_z$) amplitudes are intrinsically coupled. For remote regions of filling, the system prefers extended $s$-wave gaps in the singlet channel and $f$-wave gaps in the triplet channel (left inset). For fillings closer to zero, two degenerate solutions with $d$-wave symmetry in the $\psi$ component and $p$-wave symmetry in the $d_z$ component are found (right inset, two degenerate solutions).}
    \label{fig:sc-phases}
\end{figure}
%----------------

Additionally, for mixed-parity SC, the singlet and triplet components may describe pairing of different length scales, such that, e.g., an extended $s$-wave ($s'$) in $\psi$ can be combined with an extended $f$-wave ($f'$) in $d_z$ (as long as the two transform in the same representation). Therefore, we distinguish the mixed-parity SC states by their irreducible representations of the $C_{3v}$ symmetry group. We find that the SC phase diagram (cf.~\cref{fig:sc-phases}) is mostly governed by instabilities transforming in the $A_1$ or $E$ representations, which we label as $s/f$- and $d/p$-wave, respectively. \footnote{Note that the symmetry is enhanced to $C_{6v}$ \append{(equivalent to $D_{6h}$ due to the inherent two-dimensional nature of our model)} for $\varphi=0$, where we find pairing instabilities in the $E_2$ ($d$-wave, \append{$E_{2g}$}), $B_1$ ($f$-wave, \append{$B_{1u}$}) and $A_2$ ($i$-wave, \append{$A_{2g}$}) representations.}
To resolve the competition between superconducting instabilities, we plot the logarithm of the ratio of $d/p$-wave and $s/f$-wave amplitudes that the linearized gap equation provides as a continuous color-map. 
The two insets show examples of $d/p$-wave symmetric (right inset) and $s/f$-wave symmetric gap functions in the singlet-triplet basis. The spin-resolved gap functions on the Fermi surfaces are shown in the supplemental material~\cite{supplement}.
For all instabilities with a dominant (two-fold degenerate) $d/p$-wave instability, the free energy in a subsequent mean-field decoupling is minimized by a chiral $d+id$/$p+ip$-wave superposition of order parameters as it allows for a fully gapped Fermi surface. In the $SU(2)$ symmetric case, an $i$-wave symmetric gap function is supported in a narrow filling window close to the VHS~\cite{gneist2022competing} highlighted with green color in \cref{fig:sc-phases}.

Interestingly, for most parts of the phase diagram in \cref{fig:sc-phases}, large filling values of $|\nu|\gtrsim1/2$ support $s/f$-wave SC, whereas for small values $|\nu|\lesssim1/2$, $d/p$-wave SC is favored. The DW phases in \cref{fig:dw-phases}, on the other hand, have no clear dependence solely on $\nu$. For example, there are points of dominant in-plane spiral order at $\nu\approx-1/2$ and $\varphi\approx3\pi/8$ as well as at $\nu\approx0$ and $\varphi\approx\pi/4$ [purple in \cref{fig:dw-phases}~\captionlabel{a} and green in \cref{fig:dw-phases}~\captionlabel{b}]. The adjacent superconducting domes are of manifestly different paring symmetry, e.g., $s/f$-wave in the former and $d/p$-wave in the latter case (cf.~\cref{fig:sc-phases}).
These observations shed light on the mechanism responsible for the type of SC order: The data suggest that the precise spin and momentum structure of the dominant spin/charge fluctuations is \emph{irrelevant} as long as it is present and instead, the topology of the Fermi surfaces is responsible for the different symmetries of SC order parameter found, e.g., small pockets around $K,K'$ vs large closed lines around $\Gamma$.
The extended nature of the superconducting
instabilities ($p/d$-wave: nearest neighbors, $s/f$-wave: next-nearest neighbors) indicates that DW fluctuations with $\bvec q\neq0$ represent the pairing glue. This statement is supported by the observation that for $\varphi\lesssim\pi/2$, the DW transfer momentum is intra-VHS, i.e. $\bar{\bvec q}=\Gamma$, and SC is suppressed.
Finally, we note that at  $\varphi=\pi/3$
an additional peak at $\bvec q=K^{(\prime)}$ appears in the pairing susceptibility, indicative of enhanced pair-density-wave correlations, which were also reported recently in Ref.~\onlinecite{wu2022pair}.

\paragraph{Discussion. ---}
In this work we calculate the two-particle interaction vertex $\Gamma^{(4)}$ within the FRG to study the electronic phase diagram of a spin-orbit coupled moir\'e Hubbard model on the triangular lattice. In the group of twisted bilayer TMDs, this model is believed to have various experimental realizations through different AA-stacked homo-bilayer systems. Even more so, recent measurements show that correlated insulating and possible superconducting states are in fact realized in twisted \WSe~\cite{wang2020correlated,ghiotto2021quantum}. Our work offers an unbiased characterization of competing electronic correlations in twisted \WSe. As a result of our large-scale simulations, we provide the FRG phase diagram as a function of filling $\nu$ and displacement field $\varphi$ in the intermediate coupling regime ($U=6t$). We firmly establish a beyond mean-field characterization of intricate density-wave orderings close to the van-Hove singularity of the system. Furthermore, the FRG reveals pairing instabilities mediated by spin and charge fluctuations so that the wide variety of DW phases is complemented by a relatively large area of the phase diagram where superconducting correlations dominate. For nonzero displacement field, the SC orderings can be divided into $d/p$-wave order (including higher harmonics) for weak doping and $s/f$-wave order for strong doping. While both order parameters are unconventional in nature and caused by spin/charge fluctuations, the $s/f$-wave is nodal and the $d/p$-wave chiral ($d+id/p+ip$). Thus, we propose spectroscopy experiments on \WSe{} to verify the transition of a nodal to a chiral (fully gapped) SC order. Time-reversal symmetry breaking in the chiral state can also be detected via Kerr rotation~\cite{KAPITULNIK2015151} or muon spin relaxation~\cite{PhysRevB.89.020502}. We also note that the interplay with other nearby states can alternatively yield nematic superconductivity~\cite{PhysRevB.99.144507,PhysRevB.101.224513}, which can be detected by spatial anisotropies~\cite{Yonezawa2017,Shen2017,doi:10.1126/science.abc2836,PhysRevX.7.011009}.

In future works, we are aiming towards extending our studies on non-$SU(2)$ and multi-orbital moir\'e systems with band structures and interactions closely  motivated by materials. This includes, but is not limited to, systematic studies of longer range and cRPA-dressed interactions as an input to the FRG. Furthermore, band structures may be directly fitted to \emph{ab-initio} results~\cite{klebl2022moire} or generated with Wannierization, paving the road for high-throughput studies of competing orders in two-dimensional (moir\'e) materials.
\medskip

\begin{acknowledgments}
We thank J.~Beyer, J.~Cano, J.~Hauck, A.~Leonhardt, A.~Millis, A.~Pasupathy, A.~Schnyder, T. Sch\"afer and J.~Zang for useful discussions. We  acknowledge  funding by the Deutsche Forschungsgemeinschaft (DFG, German Research Foundation) under RTG 1995, within the Priority Program SPP 2244 ``2DMP'' and under Germany's Excellence Strategy - Cluster of Excellence Matter and Light for Quantum Computing (ML4Q) EXC 2004/1 - 390534769. We acknowledge computational resources provided by the Max Planck Computing and Data Facility and RWTH Aachen University under project number rwth0716. This work was supported by the Max Planck-New York City Center for Nonequilibrium Quantum Phenomena.
MMS acknowledges support through the DFG Heisenberg programme (project id 452976698) and SFB 1238 (project C02, project id 277146847).
\end{acknowledgments}
\medskip

\emph{During the final preparation of this manuscript,  Ref.~\onlinecite{wu2022pair} appeared providing similar conclusions where applicable.}

\bibliography{references}

\begin{thebibliography}{71}%
\makeatletter
\providecommand \@ifxundefined [1]{%
 \@ifx{#1\undefined}
}%
\providecommand \@ifnum [1]{%
 \ifnum #1\expandafter \@firstoftwo
 \else \expandafter \@secondoftwo
 \fi
}%
\providecommand \@ifx [1]{%
 \ifx #1\expandafter \@firstoftwo
 \else \expandafter \@secondoftwo
 \fi
}%
\providecommand \natexlab [1]{#1}%
\providecommand \enquote  [1]{``#1''}%
\providecommand \bibnamefont  [1]{#1}%
\providecommand \bibfnamefont [1]{#1}%
\providecommand \citenamefont [1]{#1}%
\providecommand \href@noop [0]{\@secondoftwo}%
\providecommand \href [0]{\begingroup \@sanitize@url \@href}%
\providecommand \@href[1]{\@@startlink{#1}\@@href}%
\providecommand \@@href[1]{\endgroup#1\@@endlink}%
\providecommand \@sanitize@url [0]{\catcode `\\12\catcode `\$12\catcode
  `\&12\catcode `\#12\catcode `\^12\catcode `\_12\catcode `\%12\relax}%
\providecommand \@@startlink[1]{}%
\providecommand \@@endlink[0]{}%
\providecommand \url  [0]{\begingroup\@sanitize@url \@url }%
\providecommand \@url [1]{\endgroup\@href {#1}{\urlprefix }}%
\providecommand \urlprefix  [0]{URL }%
\providecommand \Eprint [0]{\href }%
\providecommand \doibase [0]{https://doi.org/}%
\providecommand \selectlanguage [0]{\@gobble}%
\providecommand \bibinfo  [0]{\@secondoftwo}%
\providecommand \bibfield  [0]{\@secondoftwo}%
\providecommand \translation [1]{[#1]}%
\providecommand \BibitemOpen [0]{}%
\providecommand \bibitemStop [0]{}%
\providecommand \bibitemNoStop [0]{.\EOS\space}%
\providecommand \EOS [0]{\spacefactor3000\relax}%
\providecommand \BibitemShut  [1]{\csname bibitem#1\endcsname}%
\let\auto@bib@innerbib\@empty
%</preamble>
\bibitem [{\citenamefont {Kennes}\ \emph {et~al.}(2021)\citenamefont {Kennes},
  \citenamefont {Claassen}, \citenamefont {Xian}, \citenamefont {Georges},
  \citenamefont {Millis}, \citenamefont {Hone}, \citenamefont {Dean},
  \citenamefont {Basov}, \citenamefont {Pasupathy},\ and\ \citenamefont
  {Rubio}}]{moiresim}%
  \BibitemOpen
  \bibfield  {author} {\bibinfo {author} {\bibfnamefont {D.~M.}\ \bibnamefont
  {Kennes}}, \bibinfo {author} {\bibfnamefont {M.}~\bibnamefont {Claassen}},
  \bibinfo {author} {\bibfnamefont {L.}~\bibnamefont {Xian}}, \bibinfo {author}
  {\bibfnamefont {A.}~\bibnamefont {Georges}}, \bibinfo {author} {\bibfnamefont
  {A.~J.}\ \bibnamefont {Millis}}, \bibinfo {author} {\bibfnamefont
  {J.}~\bibnamefont {Hone}}, \bibinfo {author} {\bibfnamefont {C.~R.}\
  \bibnamefont {Dean}}, \bibinfo {author} {\bibfnamefont {D.~N.}\ \bibnamefont
  {Basov}}, \bibinfo {author} {\bibfnamefont {A.}~\bibnamefont {Pasupathy}},\
  and\ \bibinfo {author} {\bibfnamefont {A.}~\bibnamefont {Rubio}},\ }\bibfield
   {title} {\bibinfo {title} {Moir\'e heterostructures: a condensed matter
  quantum simulator},\ }\href@noop {} {\bibfield  {journal} {\bibinfo
  {journal} {Nat. Phys.}\ }\textbf {\bibinfo {volume} {17}},\ \bibinfo {pages}
  {155–163} (\bibinfo {year} {2021})}\BibitemShut {NoStop}%
\bibitem [{\citenamefont {Cao}\ \emph {et~al.}(2018{\natexlab{a}})\citenamefont
  {Cao}, \citenamefont {Fatemi}, \citenamefont {Fang}, \citenamefont
  {Watanabe}, \citenamefont {Taniguchi}, \citenamefont {Kaxiras},\ and\
  \citenamefont {Jarillo-Herrero}}]{cao2018unconventional}%
  \BibitemOpen
  \bibfield  {author} {\bibinfo {author} {\bibfnamefont {Y.}~\bibnamefont
  {Cao}}, \bibinfo {author} {\bibfnamefont {V.}~\bibnamefont {Fatemi}},
  \bibinfo {author} {\bibfnamefont {S.}~\bibnamefont {Fang}}, \bibinfo {author}
  {\bibfnamefont {K.}~\bibnamefont {Watanabe}}, \bibinfo {author}
  {\bibfnamefont {T.}~\bibnamefont {Taniguchi}}, \bibinfo {author}
  {\bibfnamefont {E.}~\bibnamefont {Kaxiras}},\ and\ \bibinfo {author}
  {\bibfnamefont {P.}~\bibnamefont {Jarillo-Herrero}},\ }\bibfield  {title}
  {\bibinfo {title} {Unconventional superconductivity in magic-angle graphene
  superlattices},\ }\href@noop {} {\bibfield  {journal} {\bibinfo  {journal}
  {Nature}\ }\textbf {\bibinfo {volume} {556}},\ \bibinfo {pages} {43}
  (\bibinfo {year} {2018}{\natexlab{a}})}\BibitemShut {NoStop}%
\bibitem [{\citenamefont {Cao}\ \emph {et~al.}(2018{\natexlab{b}})\citenamefont
  {Cao}, \citenamefont {Fatemi}, \citenamefont {Demir}, \citenamefont {Fang},
  \citenamefont {Tomarken}, \citenamefont {Luo}, \citenamefont
  {Sanchez-Yamagishi}, \citenamefont {Watanabe}, \citenamefont {Taniguchi},
  \citenamefont {Kaxiras}, \citenamefont {Ashoori},\ and\ \citenamefont
  {Jarillo-Herrero}}]{cao2018mott}%
  \BibitemOpen
  \bibfield  {author} {\bibinfo {author} {\bibfnamefont {Y.}~\bibnamefont
  {Cao}}, \bibinfo {author} {\bibfnamefont {V.}~\bibnamefont {Fatemi}},
  \bibinfo {author} {\bibfnamefont {A.}~\bibnamefont {Demir}}, \bibinfo
  {author} {\bibfnamefont {S.}~\bibnamefont {Fang}}, \bibinfo {author}
  {\bibfnamefont {S.~L.}\ \bibnamefont {Tomarken}}, \bibinfo {author}
  {\bibfnamefont {J.~Y.}\ \bibnamefont {Luo}}, \bibinfo {author} {\bibfnamefont
  {J.~D.}\ \bibnamefont {Sanchez-Yamagishi}}, \bibinfo {author} {\bibfnamefont
  {K.}~\bibnamefont {Watanabe}}, \bibinfo {author} {\bibfnamefont
  {T.}~\bibnamefont {Taniguchi}}, \bibinfo {author} {\bibfnamefont
  {E.}~\bibnamefont {Kaxiras}}, \bibinfo {author} {\bibfnamefont {R.~C.}\
  \bibnamefont {Ashoori}},\ and\ \bibinfo {author} {\bibfnamefont
  {P.}~\bibnamefont {Jarillo-Herrero}},\ }\bibfield  {title} {\bibinfo {title}
  {Correlated insulator behaviour at half-filling in magic-angle graphene
  superlattices},\ }\href@noop {} {\bibfield  {journal} {\bibinfo  {journal}
  {Nature}\ }\textbf {\bibinfo {volume} {556}},\ \bibinfo {pages} {80}
  (\bibinfo {year} {2018}{\natexlab{b}})}\BibitemShut {NoStop}%
\bibitem [{\citenamefont {Lu}\ \emph {et~al.}(2019)\citenamefont {Lu},
  \citenamefont {Stepanov}, \citenamefont {Yang}, \citenamefont {Xie},
  \citenamefont {Aamir}, \citenamefont {Das}, \citenamefont {Urgell},
  \citenamefont {Watanabe}, \citenamefont {Taniguchi}, \citenamefont {Zhang},
  \citenamefont {Bachtold}, \citenamefont {MacDonald},\ and\ \citenamefont
  {Efetov}}]{lu2019superconductors}%
  \BibitemOpen
  \bibfield  {author} {\bibinfo {author} {\bibfnamefont {X.}~\bibnamefont
  {Lu}}, \bibinfo {author} {\bibfnamefont {P.}~\bibnamefont {Stepanov}},
  \bibinfo {author} {\bibfnamefont {W.}~\bibnamefont {Yang}}, \bibinfo {author}
  {\bibfnamefont {M.}~\bibnamefont {Xie}}, \bibinfo {author} {\bibfnamefont
  {M.~A.}\ \bibnamefont {Aamir}}, \bibinfo {author} {\bibfnamefont
  {I.}~\bibnamefont {Das}}, \bibinfo {author} {\bibfnamefont {C.}~\bibnamefont
  {Urgell}}, \bibinfo {author} {\bibfnamefont {K.}~\bibnamefont {Watanabe}},
  \bibinfo {author} {\bibfnamefont {T.}~\bibnamefont {Taniguchi}}, \bibinfo
  {author} {\bibfnamefont {G.}~\bibnamefont {Zhang}}, \bibinfo {author}
  {\bibfnamefont {A.}~\bibnamefont {Bachtold}}, \bibinfo {author}
  {\bibfnamefont {A.~H.}\ \bibnamefont {MacDonald}},\ and\ \bibinfo {author}
  {\bibfnamefont {D.~K.}\ \bibnamefont {Efetov}},\ }\bibfield  {title}
  {\bibinfo {title} {Superconductors, orbital magnets and correlated states in
  magic-angle bilayer graphene},\ }\href@noop {} {\bibfield  {journal}
  {\bibinfo  {journal} {Nature}\ }\textbf {\bibinfo {volume} {574}},\ \bibinfo
  {pages} {653} (\bibinfo {year} {2019})}\BibitemShut {NoStop}%
\bibitem [{\citenamefont {Cao}\ \emph {et~al.}(2020{\natexlab{a}})\citenamefont
  {Cao}, \citenamefont {Chowdhury}, \citenamefont {Rodan-Legrain},
  \citenamefont {Rubies-Bigord\`a}, \citenamefont {Watanabe}, \citenamefont
  {Taniguchi}, \citenamefont {Senthil},\ and\ \citenamefont
  {Jarillo-Herrero}}]{Cao2020strange}%
  \BibitemOpen
  \bibfield  {author} {\bibinfo {author} {\bibfnamefont {Y.}~\bibnamefont
  {Cao}}, \bibinfo {author} {\bibfnamefont {D.}~\bibnamefont {Chowdhury}},
  \bibinfo {author} {\bibfnamefont {D.}~\bibnamefont {Rodan-Legrain}}, \bibinfo
  {author} {\bibfnamefont {O.}~\bibnamefont {Rubies-Bigord\`a}}, \bibinfo
  {author} {\bibfnamefont {K.}~\bibnamefont {Watanabe}}, \bibinfo {author}
  {\bibfnamefont {T.}~\bibnamefont {Taniguchi}}, \bibinfo {author}
  {\bibfnamefont {T.}~\bibnamefont {Senthil}},\ and\ \bibinfo {author}
  {\bibfnamefont {P.}~\bibnamefont {Jarillo-Herrero}},\ }\bibfield  {title}
  {\bibinfo {title} {Strange metal in magic-angle graphene with near planckian
  dissipation},\ }\href@noop {} {\bibfield  {journal} {\bibinfo  {journal}
  {Phys. Rev. Lett.}\ }\textbf {\bibinfo {volume} {124}},\ \bibinfo {pages}
  {076801} (\bibinfo {year} {2020}{\natexlab{a}})}\BibitemShut {NoStop}%
\bibitem [{\citenamefont {Polshyn}\ \emph {et~al.}(2019)\citenamefont
  {Polshyn}, \citenamefont {Yankowitz}, \citenamefont {Chen}, \citenamefont
  {Zhang}, \citenamefont {Watanabe}, \citenamefont {Taniguchi}, \citenamefont
  {Dean},\ and\ \citenamefont {Young}}]{Polshyn2019}%
  \BibitemOpen
  \bibfield  {author} {\bibinfo {author} {\bibfnamefont {H.}~\bibnamefont
  {Polshyn}}, \bibinfo {author} {\bibfnamefont {M.}~\bibnamefont {Yankowitz}},
  \bibinfo {author} {\bibfnamefont {S.}~\bibnamefont {Chen}}, \bibinfo {author}
  {\bibfnamefont {Y.}~\bibnamefont {Zhang}}, \bibinfo {author} {\bibfnamefont
  {K.}~\bibnamefont {Watanabe}}, \bibinfo {author} {\bibfnamefont
  {T.}~\bibnamefont {Taniguchi}}, \bibinfo {author} {\bibfnamefont {C.~R.}\
  \bibnamefont {Dean}},\ and\ \bibinfo {author} {\bibfnamefont {A.~F.}\
  \bibnamefont {Young}},\ }\bibfield  {title} {\bibinfo {title} {Large
  linear-in-temperature resistivity in twisted bilayer graphene},\ }\href@noop
  {} {\bibfield  {journal} {\bibinfo  {journal} {Nat. Phys.}\ }\textbf
  {\bibinfo {volume} {15}},\ \bibinfo {pages} {1011} (\bibinfo {year}
  {2019})}\BibitemShut {NoStop}%
\bibitem [{\citenamefont {Yankowitz}\ \emph {et~al.}(2019)\citenamefont
  {Yankowitz}, \citenamefont {Chen}, \citenamefont {Polshyn}, \citenamefont
  {Zhang}, \citenamefont {Watanabe}, \citenamefont {Taniguchi}, \citenamefont
  {Graf}, \citenamefont {Young},\ and\ \citenamefont
  {Dean}}]{yankowitz2019tuning}%
  \BibitemOpen
  \bibfield  {author} {\bibinfo {author} {\bibfnamefont {M.}~\bibnamefont
  {Yankowitz}}, \bibinfo {author} {\bibfnamefont {S.}~\bibnamefont {Chen}},
  \bibinfo {author} {\bibfnamefont {H.}~\bibnamefont {Polshyn}}, \bibinfo
  {author} {\bibfnamefont {Y.}~\bibnamefont {Zhang}}, \bibinfo {author}
  {\bibfnamefont {K.}~\bibnamefont {Watanabe}}, \bibinfo {author}
  {\bibfnamefont {T.}~\bibnamefont {Taniguchi}}, \bibinfo {author}
  {\bibfnamefont {D.}~\bibnamefont {Graf}}, \bibinfo {author} {\bibfnamefont
  {A.~F.}\ \bibnamefont {Young}},\ and\ \bibinfo {author} {\bibfnamefont
  {C.~R.}\ \bibnamefont {Dean}},\ }\bibfield  {title} {\bibinfo {title} {Tuning
  superconductivity in twisted bilayer graphene},\ }\href@noop {} {\bibfield
  {journal} {\bibinfo  {journal} {Science}\ }\textbf {\bibinfo {volume}
  {363}},\ \bibinfo {pages} {1059} (\bibinfo {year} {2019})}\BibitemShut
  {NoStop}%
\bibitem [{\citenamefont {Liu}\ \emph {et~al.}(2021)\citenamefont {Liu},
  \citenamefont {Wang}, \citenamefont {Watanabe}, \citenamefont {Taniguchi},
  \citenamefont {Vafek},\ and\ \citenamefont {Li}}]{liu2021tuning}%
  \BibitemOpen
  \bibfield  {author} {\bibinfo {author} {\bibfnamefont {X.}~\bibnamefont
  {Liu}}, \bibinfo {author} {\bibfnamefont {Z.}~\bibnamefont {Wang}}, \bibinfo
  {author} {\bibfnamefont {K.}~\bibnamefont {Watanabe}}, \bibinfo {author}
  {\bibfnamefont {T.}~\bibnamefont {Taniguchi}}, \bibinfo {author}
  {\bibfnamefont {O.}~\bibnamefont {Vafek}},\ and\ \bibinfo {author}
  {\bibfnamefont {J.}~\bibnamefont {Li}},\ }\bibfield  {title} {\bibinfo
  {title} {Tuning electron correlation in magic-angle twisted bilayer graphene
  using coulomb screening},\ }\href@noop {} {\bibfield  {journal} {\bibinfo
  {journal} {Science}\ }\textbf {\bibinfo {volume} {371}},\ \bibinfo {pages}
  {1261} (\bibinfo {year} {2021})}\BibitemShut {NoStop}%
\bibitem [{\citenamefont {Stepanov}\ \emph {et~al.}(2020)\citenamefont
  {Stepanov}, \citenamefont {Das}, \citenamefont {Lu}, \citenamefont
  {Fahimniya}, \citenamefont {Watanabe}, \citenamefont {Taniguchi},
  \citenamefont {Koppens}, \citenamefont {Lischner}, \citenamefont {Levitov},\
  and\ \citenamefont {Efetov}}]{stepanov2020untying}%
  \BibitemOpen
  \bibfield  {author} {\bibinfo {author} {\bibfnamefont {P.}~\bibnamefont
  {Stepanov}}, \bibinfo {author} {\bibfnamefont {I.}~\bibnamefont {Das}},
  \bibinfo {author} {\bibfnamefont {X.}~\bibnamefont {Lu}}, \bibinfo {author}
  {\bibfnamefont {A.}~\bibnamefont {Fahimniya}}, \bibinfo {author}
  {\bibfnamefont {K.}~\bibnamefont {Watanabe}}, \bibinfo {author}
  {\bibfnamefont {T.}~\bibnamefont {Taniguchi}}, \bibinfo {author}
  {\bibfnamefont {F.~H.}\ \bibnamefont {Koppens}}, \bibinfo {author}
  {\bibfnamefont {J.}~\bibnamefont {Lischner}}, \bibinfo {author}
  {\bibfnamefont {L.}~\bibnamefont {Levitov}},\ and\ \bibinfo {author}
  {\bibfnamefont {D.~K.}\ \bibnamefont {Efetov}},\ }\bibfield  {title}
  {\bibinfo {title} {Untying the insulating and superconducting orders in
  magic-angle graphene},\ }\href@noop {} {\bibfield  {journal} {\bibinfo
  {journal} {Nature}\ }\textbf {\bibinfo {volume} {583}},\ \bibinfo {pages}
  {375} (\bibinfo {year} {2020})}\BibitemShut {NoStop}%
\bibitem [{\citenamefont {Arora}\ \emph {et~al.}(2020)\citenamefont {Arora},
  \citenamefont {Polski}, \citenamefont {Zhang}, \citenamefont {Thomson},
  \citenamefont {Choi}, \citenamefont {Kim}, \citenamefont {Lin}, \citenamefont
  {Wilson}, \citenamefont {Xu}, \citenamefont {Chu} \emph
  {et~al.}}]{arora2020superconductivity}%
  \BibitemOpen
  \bibfield  {author} {\bibinfo {author} {\bibfnamefont {H.~S.}\ \bibnamefont
  {Arora}}, \bibinfo {author} {\bibfnamefont {R.}~\bibnamefont {Polski}},
  \bibinfo {author} {\bibfnamefont {Y.}~\bibnamefont {Zhang}}, \bibinfo
  {author} {\bibfnamefont {A.}~\bibnamefont {Thomson}}, \bibinfo {author}
  {\bibfnamefont {Y.}~\bibnamefont {Choi}}, \bibinfo {author} {\bibfnamefont
  {H.}~\bibnamefont {Kim}}, \bibinfo {author} {\bibfnamefont {Z.}~\bibnamefont
  {Lin}}, \bibinfo {author} {\bibfnamefont {I.~Z.}\ \bibnamefont {Wilson}},
  \bibinfo {author} {\bibfnamefont {X.}~\bibnamefont {Xu}}, \bibinfo {author}
  {\bibfnamefont {J.-H.}\ \bibnamefont {Chu}}, \emph {et~al.},\ }\bibfield
  {title} {\bibinfo {title} {Superconductivity in metallic twisted bilayer
  graphene stabilized by wse 2},\ }\href@noop {} {\bibfield  {journal}
  {\bibinfo  {journal} {Nature}\ }\textbf {\bibinfo {volume} {583}},\ \bibinfo
  {pages} {379} (\bibinfo {year} {2020})}\BibitemShut {NoStop}%
\bibitem [{\citenamefont {Zondiner}\ \emph {et~al.}(2020)\citenamefont
  {Zondiner}, \citenamefont {Rozen}, \citenamefont {Rodan-Legrain},
  \citenamefont {Cao}, \citenamefont {Queiroz}, \citenamefont {Taniguchi},
  \citenamefont {Watanabe}, \citenamefont {Oreg}, \citenamefont {von Oppen},
  \citenamefont {Stern}, \citenamefont {Berg}, \citenamefont
  {Jarillo-Herrero},\ and\ \citenamefont {Ilani}}]{Zondiner2020}%
  \BibitemOpen
  \bibfield  {author} {\bibinfo {author} {\bibfnamefont {U.}~\bibnamefont
  {Zondiner}}, \bibinfo {author} {\bibfnamefont {A.}~\bibnamefont {Rozen}},
  \bibinfo {author} {\bibfnamefont {D.}~\bibnamefont {Rodan-Legrain}}, \bibinfo
  {author} {\bibfnamefont {Y.}~\bibnamefont {Cao}}, \bibinfo {author}
  {\bibfnamefont {R.}~\bibnamefont {Queiroz}}, \bibinfo {author} {\bibfnamefont
  {T.}~\bibnamefont {Taniguchi}}, \bibinfo {author} {\bibfnamefont
  {K.}~\bibnamefont {Watanabe}}, \bibinfo {author} {\bibfnamefont
  {Y.}~\bibnamefont {Oreg}}, \bibinfo {author} {\bibfnamefont {F.}~\bibnamefont
  {von Oppen}}, \bibinfo {author} {\bibfnamefont {A.}~\bibnamefont {Stern}},
  \bibinfo {author} {\bibfnamefont {E.}~\bibnamefont {Berg}}, \bibinfo {author}
  {\bibfnamefont {P.}~\bibnamefont {Jarillo-Herrero}},\ and\ \bibinfo {author}
  {\bibfnamefont {S.}~\bibnamefont {Ilani}},\ }\bibfield  {title} {\bibinfo
  {title} {Cascade of phase transitions and dirac revivals in magic-angle
  graphene},\ }\href@noop {} {\bibfield  {journal} {\bibinfo  {journal}
  {Nature}\ }\textbf {\bibinfo {volume} {582}},\ \bibinfo {pages} {203}
  (\bibinfo {year} {2020})}\BibitemShut {NoStop}%
\bibitem [{\citenamefont {Wong}\ \emph {et~al.}(2020)\citenamefont {Wong},
  \citenamefont {Nuckolls}, \citenamefont {Oh}, \citenamefont {Lian},
  \citenamefont {Yonglong~Xie}, \citenamefont {Watanabe}, \citenamefont
  {Taniguchi}, \citenamefont {Bernevig},\ and\ \citenamefont
  {Yazdani}}]{Wong2020}%
  \BibitemOpen
  \bibfield  {author} {\bibinfo {author} {\bibfnamefont {D.}~\bibnamefont
  {Wong}}, \bibinfo {author} {\bibfnamefont {K.~P.}\ \bibnamefont {Nuckolls}},
  \bibinfo {author} {\bibfnamefont {M.}~\bibnamefont {Oh}}, \bibinfo {author}
  {\bibfnamefont {B.}~\bibnamefont {Lian}}, \bibinfo {author} {\bibfnamefont
  {S.~J.}\ \bibnamefont {Yonglong~Xie}}, \bibinfo {author} {\bibfnamefont
  {K.}~\bibnamefont {Watanabe}}, \bibinfo {author} {\bibfnamefont
  {T.}~\bibnamefont {Taniguchi}}, \bibinfo {author} {\bibfnamefont {B.~A.}\
  \bibnamefont {Bernevig}},\ and\ \bibinfo {author} {\bibfnamefont
  {A.}~\bibnamefont {Yazdani}},\ }\bibfield  {title} {\bibinfo {title} {Cascade
  of electronic transitions in magic-angle twisted bilayer graphene},\
  }\href@noop {} {\bibfield  {journal} {\bibinfo  {journal} {Nature}\ }\textbf
  {\bibinfo {volume} {582}},\ \bibinfo {pages} {198–202} (\bibinfo {year}
  {2020})}\BibitemShut {NoStop}%
\bibitem [{\citenamefont {Xie}\ \emph {et~al.}(2019)\citenamefont {Xie},
  \citenamefont {Lian}, \citenamefont {J\"{a}ck}, \citenamefont {Liu},
  \citenamefont {Chiu}, \citenamefont {Watanabe}, \citenamefont {Taniguchi},
  \citenamefont {Bernevig},\ and\ \citenamefont
  {Yazdani}}]{Xie2019spectrosopic}%
  \BibitemOpen
  \bibfield  {author} {\bibinfo {author} {\bibfnamefont {Y.}~\bibnamefont
  {Xie}}, \bibinfo {author} {\bibfnamefont {B.}~\bibnamefont {Lian}}, \bibinfo
  {author} {\bibfnamefont {B.}~\bibnamefont {J\"{a}ck}}, \bibinfo {author}
  {\bibfnamefont {X.}~\bibnamefont {Liu}}, \bibinfo {author} {\bibfnamefont
  {C.-L.}\ \bibnamefont {Chiu}}, \bibinfo {author} {\bibfnamefont
  {K.}~\bibnamefont {Watanabe}}, \bibinfo {author} {\bibfnamefont
  {T.}~\bibnamefont {Taniguchi}}, \bibinfo {author} {\bibfnamefont {B.~A.}\
  \bibnamefont {Bernevig}},\ and\ \bibinfo {author} {\bibfnamefont
  {A.}~\bibnamefont {Yazdani}},\ }\bibfield  {title} {\bibinfo {title}
  {Spectroscopic signatures of many-body correlations in magic-angle twisted
  bilayer graphene},\ }\href@noop {} {\bibfield  {journal} {\bibinfo  {journal}
  {Nature}\ }\textbf {\bibinfo {volume} {572}},\ \bibinfo {pages} {101}
  (\bibinfo {year} {2019})}\BibitemShut {NoStop}%
\bibitem [{\citenamefont {Kerelsky}\ \emph {et~al.}(2019)\citenamefont
  {Kerelsky}, \citenamefont {McGilly}, \citenamefont {Kennes}, \citenamefont
  {Xian}, \citenamefont {Yankowitz}, \citenamefont {Chen}, \citenamefont
  {Watanabe}, \citenamefont {Taniguchi}, \citenamefont {Hone}, \citenamefont
  {Dean}, \citenamefont {Rubio},\ and\ \citenamefont
  {Pasupathy}}]{Kerelsky2019maximized}%
  \BibitemOpen
  \bibfield  {author} {\bibinfo {author} {\bibfnamefont {A.}~\bibnamefont
  {Kerelsky}}, \bibinfo {author} {\bibfnamefont {L.~J.}\ \bibnamefont
  {McGilly}}, \bibinfo {author} {\bibfnamefont {D.~M.}\ \bibnamefont {Kennes}},
  \bibinfo {author} {\bibfnamefont {L.}~\bibnamefont {Xian}}, \bibinfo {author}
  {\bibfnamefont {M.}~\bibnamefont {Yankowitz}}, \bibinfo {author}
  {\bibfnamefont {S.}~\bibnamefont {Chen}}, \bibinfo {author} {\bibfnamefont
  {K.}~\bibnamefont {Watanabe}}, \bibinfo {author} {\bibfnamefont
  {T.}~\bibnamefont {Taniguchi}}, \bibinfo {author} {\bibfnamefont
  {J.}~\bibnamefont {Hone}}, \bibinfo {author} {\bibfnamefont {C.}~\bibnamefont
  {Dean}}, \bibinfo {author} {\bibfnamefont {A.}~\bibnamefont {Rubio}},\ and\
  \bibinfo {author} {\bibfnamefont {A.~N.}\ \bibnamefont {Pasupathy}},\
  }\bibfield  {title} {\bibinfo {title} {Maximized electron interactions at the
  magic angle in twisted bilayer graphene},\ }\href@noop {} {\bibfield
  {journal} {\bibinfo  {journal} {Nature}\ }\textbf {\bibinfo {volume} {572}},\
  \bibinfo {pages} {95} (\bibinfo {year} {2019})}\BibitemShut {NoStop}%
\bibitem [{\citenamefont {Jiang}\ \emph {et~al.}(2019)\citenamefont {Jiang},
  \citenamefont {Lai}, \citenamefont {Watanabe}, \citenamefont {Taniguchi},
  \citenamefont {Haule}, \citenamefont {Mao},\ and\ \citenamefont
  {Andrei}}]{Jiang2019charge}%
  \BibitemOpen
  \bibfield  {author} {\bibinfo {author} {\bibfnamefont {Y.}~\bibnamefont
  {Jiang}}, \bibinfo {author} {\bibfnamefont {X.}~\bibnamefont {Lai}}, \bibinfo
  {author} {\bibfnamefont {K.}~\bibnamefont {Watanabe}}, \bibinfo {author}
  {\bibfnamefont {T.}~\bibnamefont {Taniguchi}}, \bibinfo {author}
  {\bibfnamefont {K.}~\bibnamefont {Haule}}, \bibinfo {author} {\bibfnamefont
  {J.}~\bibnamefont {Mao}},\ and\ \bibinfo {author} {\bibfnamefont {E.~Y.}\
  \bibnamefont {Andrei}},\ }\bibfield  {title} {\bibinfo {title} {Charge order
  and broken rotational symmetry in magic-angle twisted bilayer graphene},\
  }\href@noop {} {\bibfield  {journal} {\bibinfo  {journal} {Nature}\ }\textbf
  {\bibinfo {volume} {573}},\ \bibinfo {pages} {91} (\bibinfo {year}
  {2019})}\BibitemShut {NoStop}%
\bibitem [{\citenamefont {Choi}\ \emph {et~al.}(2019)\citenamefont {Choi},
  \citenamefont {Kemmer}, \citenamefont {Peng}, \citenamefont {Thomson},
  \citenamefont {Arora}, \citenamefont {Polski}, \citenamefont {Zhang},
  \citenamefont {Ren}, \citenamefont {Alicea}, \citenamefont {Refael},
  \citenamefont {von Oppen}, \citenamefont {Watanabe}, \citenamefont
  {Taniguchi},\ and\ \citenamefont {Nadj-Perge}}]{Choi2019correlations}%
  \BibitemOpen
  \bibfield  {author} {\bibinfo {author} {\bibfnamefont {Y.}~\bibnamefont
  {Choi}}, \bibinfo {author} {\bibfnamefont {J.}~\bibnamefont {Kemmer}},
  \bibinfo {author} {\bibfnamefont {Y.}~\bibnamefont {Peng}}, \bibinfo {author}
  {\bibfnamefont {A.}~\bibnamefont {Thomson}}, \bibinfo {author} {\bibfnamefont
  {H.}~\bibnamefont {Arora}}, \bibinfo {author} {\bibfnamefont
  {R.}~\bibnamefont {Polski}}, \bibinfo {author} {\bibfnamefont
  {Y.}~\bibnamefont {Zhang}}, \bibinfo {author} {\bibfnamefont
  {H.}~\bibnamefont {Ren}}, \bibinfo {author} {\bibfnamefont {J.}~\bibnamefont
  {Alicea}}, \bibinfo {author} {\bibfnamefont {G.}~\bibnamefont {Refael}},
  \bibinfo {author} {\bibfnamefont {F.}~\bibnamefont {von Oppen}}, \bibinfo
  {author} {\bibfnamefont {K.}~\bibnamefont {Watanabe}}, \bibinfo {author}
  {\bibfnamefont {T.}~\bibnamefont {Taniguchi}},\ and\ \bibinfo {author}
  {\bibfnamefont {S.}~\bibnamefont {Nadj-Perge}},\ }\bibfield  {title}
  {\bibinfo {title} {Electronic correlations in twisted bilayer graphene near
  the magic angle},\ }\href@noop {} {\bibfield  {journal} {\bibinfo  {journal}
  {Nat. Phys.}\ }\textbf {\bibinfo {volume} {15}},\ \bibinfo {pages} {1174}
  (\bibinfo {year} {2019})}\BibitemShut {NoStop}%
\bibitem [{\citenamefont {Cao}\ \emph {et~al.}(2020{\natexlab{b}})\citenamefont
  {Cao}, \citenamefont {Rodan-Legrain}, \citenamefont {Park}, \citenamefont
  {Yuan}, \citenamefont {Watanabe}, \citenamefont {Taniguchi}, \citenamefont
  {Fernandes}, \citenamefont {Fu},\ and\ \citenamefont
  {Jarillo-Herrero}}]{cao2020nematicity}%
  \BibitemOpen
  \bibfield  {author} {\bibinfo {author} {\bibfnamefont {Y.}~\bibnamefont
  {Cao}}, \bibinfo {author} {\bibfnamefont {D.}~\bibnamefont {Rodan-Legrain}},
  \bibinfo {author} {\bibfnamefont {J.~M.}\ \bibnamefont {Park}}, \bibinfo
  {author} {\bibfnamefont {F.~N.}\ \bibnamefont {Yuan}}, \bibinfo {author}
  {\bibfnamefont {K.}~\bibnamefont {Watanabe}}, \bibinfo {author}
  {\bibfnamefont {T.}~\bibnamefont {Taniguchi}}, \bibinfo {author}
  {\bibfnamefont {R.~M.}\ \bibnamefont {Fernandes}}, \bibinfo {author}
  {\bibfnamefont {L.}~\bibnamefont {Fu}},\ and\ \bibinfo {author}
  {\bibfnamefont {P.}~\bibnamefont {Jarillo-Herrero}},\ }\bibfield  {title}
  {\bibinfo {title} {Nematicity and competing orders in superconducting
  magic-angle graphene},\ }\href@noop {} {\bibfield  {journal} {\bibinfo
  {journal} {arXiv preprint arXiv:2004.04148}\ } (\bibinfo {year}
  {2020}{\natexlab{b}})}\BibitemShut {NoStop}%
\bibitem [{\citenamefont {Burg}\ \emph
  {et~al.}(2019{\natexlab{a}})\citenamefont {Burg}, \citenamefont {Zhu},
  \citenamefont {Taniguchi}, \citenamefont {Watanabe}, \citenamefont
  {MacDonald},\ and\ \citenamefont {Tutuc}}]{burg2019correlated}%
  \BibitemOpen
  \bibfield  {author} {\bibinfo {author} {\bibfnamefont {G.~W.}\ \bibnamefont
  {Burg}}, \bibinfo {author} {\bibfnamefont {J.}~\bibnamefont {Zhu}}, \bibinfo
  {author} {\bibfnamefont {T.}~\bibnamefont {Taniguchi}}, \bibinfo {author}
  {\bibfnamefont {K.}~\bibnamefont {Watanabe}}, \bibinfo {author}
  {\bibfnamefont {A.~H.}\ \bibnamefont {MacDonald}},\ and\ \bibinfo {author}
  {\bibfnamefont {E.}~\bibnamefont {Tutuc}},\ }\bibfield  {title} {\bibinfo
  {title} {Correlated insulating states in twisted double bilayer graphene},\
  }\href@noop {} {\bibfield  {journal} {\bibinfo  {journal} {Physical review
  letters}\ }\textbf {\bibinfo {volume} {123}},\ \bibinfo {pages} {197702}
  (\bibinfo {year} {2019}{\natexlab{a}})}\BibitemShut {NoStop}%
\bibitem [{\citenamefont {Park}\ \emph
  {et~al.}(2021{\natexlab{a}})\citenamefont {Park}, \citenamefont {Cao},
  \citenamefont {Watanabe}, \citenamefont {Taniguchi},\ and\ \citenamefont
  {Jarillo-Herrero}}]{Park2021}%
  \BibitemOpen
  \bibfield  {author} {\bibinfo {author} {\bibfnamefont {J.~M.}\ \bibnamefont
  {Park}}, \bibinfo {author} {\bibfnamefont {Y.}~\bibnamefont {Cao}}, \bibinfo
  {author} {\bibfnamefont {K.}~\bibnamefont {Watanabe}}, \bibinfo {author}
  {\bibfnamefont {T.}~\bibnamefont {Taniguchi}},\ and\ \bibinfo {author}
  {\bibfnamefont {P.}~\bibnamefont {Jarillo-Herrero}},\ }\bibfield  {title}
  {\bibinfo {title} {Tunable strongly coupled superconductivity in magic-angle
  twisted trilayer graphene},\ }\href
  {https://doi.org/10.1038/s41586-021-03192-0} {\bibfield  {journal} {\bibinfo
  {journal} {Nature}\ }\textbf {\bibinfo {volume} {590}},\ \bibinfo {pages}
  {249} (\bibinfo {year} {2021}{\natexlab{a}})}\BibitemShut {NoStop}%
\bibitem [{\citenamefont {Cao}\ \emph {et~al.}(2021{\natexlab{a}})\citenamefont
  {Cao}, \citenamefont {Park}, \citenamefont {Watanabe}, \citenamefont
  {Taniguchi},\ and\ \citenamefont {Jarillo-Herrero}}]{cao2021large}%
  \BibitemOpen
  \bibfield  {author} {\bibinfo {author} {\bibfnamefont {Y.}~\bibnamefont
  {Cao}}, \bibinfo {author} {\bibfnamefont {J.~M.}\ \bibnamefont {Park}},
  \bibinfo {author} {\bibfnamefont {K.}~\bibnamefont {Watanabe}}, \bibinfo
  {author} {\bibfnamefont {T.}~\bibnamefont {Taniguchi}},\ and\ \bibinfo
  {author} {\bibfnamefont {P.}~\bibnamefont {Jarillo-Herrero}},\ }\href@noop {}
  {\bibinfo {title} {Large pauli limit violation and reentrant
  superconductivity in magic-angle twisted trilayer graphene}} (\bibinfo {year}
  {2021}{\natexlab{a}}),\ \Eprint {https://arxiv.org/abs/2103.12083}
  {arXiv:2103.12083 [cond-mat.mes-hall]} \BibitemShut {NoStop}%
\bibitem [{\citenamefont {Hao}\ \emph {et~al.}(2021)\citenamefont {Hao},
  \citenamefont {Zimmerman}, \citenamefont {Ledwith}, \citenamefont {Khalaf},
  \citenamefont {Najafabadi}, \citenamefont {Watanabe}, \citenamefont
  {Taniguchi}, \citenamefont {Vishwanath},\ and\ \citenamefont
  {Kim}}]{hao2021electric}%
  \BibitemOpen
  \bibfield  {author} {\bibinfo {author} {\bibfnamefont {Z.}~\bibnamefont
  {Hao}}, \bibinfo {author} {\bibfnamefont {A.}~\bibnamefont {Zimmerman}},
  \bibinfo {author} {\bibfnamefont {P.}~\bibnamefont {Ledwith}}, \bibinfo
  {author} {\bibfnamefont {E.}~\bibnamefont {Khalaf}}, \bibinfo {author}
  {\bibfnamefont {D.~H.}\ \bibnamefont {Najafabadi}}, \bibinfo {author}
  {\bibfnamefont {K.}~\bibnamefont {Watanabe}}, \bibinfo {author}
  {\bibfnamefont {T.}~\bibnamefont {Taniguchi}}, \bibinfo {author}
  {\bibfnamefont {A.}~\bibnamefont {Vishwanath}},\ and\ \bibinfo {author}
  {\bibfnamefont {P.}~\bibnamefont {Kim}},\ }\bibfield  {title} {\bibinfo
  {title} {Electric field--tunable superconductivity in alternating-twist
  magic-angle trilayer graphene},\ }\href@noop {} {\bibfield  {journal}
  {\bibinfo  {journal} {Science}\ }\textbf {\bibinfo {volume} {371}},\ \bibinfo
  {pages} {1133} (\bibinfo {year} {2021})}\BibitemShut {NoStop}%
\bibitem [{\citenamefont {Kim}\ \emph {et~al.}(2021)\citenamefont {Kim},
  \citenamefont {Choi}, \citenamefont {Lewandowski}, \citenamefont {Thomson},
  \citenamefont {Zhang}, \citenamefont {Polski}, \citenamefont {Watanabe},
  \citenamefont {Taniguchi}, \citenamefont {Alicea},\ and\ \citenamefont
  {Nadj-Perge}}]{kim2021spectroscopic}%
  \BibitemOpen
  \bibfield  {author} {\bibinfo {author} {\bibfnamefont {H.}~\bibnamefont
  {Kim}}, \bibinfo {author} {\bibfnamefont {Y.}~\bibnamefont {Choi}}, \bibinfo
  {author} {\bibfnamefont {C.}~\bibnamefont {Lewandowski}}, \bibinfo {author}
  {\bibfnamefont {A.}~\bibnamefont {Thomson}}, \bibinfo {author} {\bibfnamefont
  {Y.}~\bibnamefont {Zhang}}, \bibinfo {author} {\bibfnamefont
  {R.}~\bibnamefont {Polski}}, \bibinfo {author} {\bibfnamefont
  {K.}~\bibnamefont {Watanabe}}, \bibinfo {author} {\bibfnamefont
  {T.}~\bibnamefont {Taniguchi}}, \bibinfo {author} {\bibfnamefont
  {J.}~\bibnamefont {Alicea}},\ and\ \bibinfo {author} {\bibfnamefont
  {S.}~\bibnamefont {Nadj-Perge}},\ }\href@noop {} {\bibinfo {title}
  {Spectroscopic signatures of strong correlations and unconventional
  superconductivity in twisted trilayer graphene}} (\bibinfo {year} {2021}),\
  \Eprint {https://arxiv.org/abs/2109.12127} {arXiv:2109.12127
  [cond-mat.mes-hall]} \BibitemShut {NoStop}%
\bibitem [{\citenamefont {Liu}\ \emph {et~al.}(2020)\citenamefont {Liu},
  \citenamefont {Hao}, \citenamefont {Khalaf}, \citenamefont {Lee},
  \citenamefont {Ronen}, \citenamefont {Yoo}, \citenamefont {Najafabadi},
  \citenamefont {Watanabe}, \citenamefont {Taniguchi}, \citenamefont
  {Vishwanath} \emph {et~al.}}]{liu2020tunable}%
  \BibitemOpen
  \bibfield  {author} {\bibinfo {author} {\bibfnamefont {X.}~\bibnamefont
  {Liu}}, \bibinfo {author} {\bibfnamefont {Z.}~\bibnamefont {Hao}}, \bibinfo
  {author} {\bibfnamefont {E.}~\bibnamefont {Khalaf}}, \bibinfo {author}
  {\bibfnamefont {J.~Y.}\ \bibnamefont {Lee}}, \bibinfo {author} {\bibfnamefont
  {Y.}~\bibnamefont {Ronen}}, \bibinfo {author} {\bibfnamefont
  {H.}~\bibnamefont {Yoo}}, \bibinfo {author} {\bibfnamefont {D.~H.}\
  \bibnamefont {Najafabadi}}, \bibinfo {author} {\bibfnamefont
  {K.}~\bibnamefont {Watanabe}}, \bibinfo {author} {\bibfnamefont
  {T.}~\bibnamefont {Taniguchi}}, \bibinfo {author} {\bibfnamefont
  {A.}~\bibnamefont {Vishwanath}}, \emph {et~al.},\ }\bibfield  {title}
  {\bibinfo {title} {Tunable spin-polarized correlated states in twisted double
  bilayer graphene},\ }\href@noop {} {\bibfield  {journal} {\bibinfo  {journal}
  {Nature}\ }\textbf {\bibinfo {volume} {583}},\ \bibinfo {pages} {221}
  (\bibinfo {year} {2020})}\BibitemShut {NoStop}%
\bibitem [{\citenamefont {Shen}\ \emph {et~al.}(2020)\citenamefont {Shen},
  \citenamefont {Chu}, \citenamefont {Wu}, \citenamefont {Li}, \citenamefont
  {Wang}, \citenamefont {Zhao}, \citenamefont {Tang}, \citenamefont {Liu},
  \citenamefont {Tian}, \citenamefont {Watanabe} \emph
  {et~al.}}]{shen2020correlated}%
  \BibitemOpen
  \bibfield  {author} {\bibinfo {author} {\bibfnamefont {C.}~\bibnamefont
  {Shen}}, \bibinfo {author} {\bibfnamefont {Y.}~\bibnamefont {Chu}}, \bibinfo
  {author} {\bibfnamefont {Q.}~\bibnamefont {Wu}}, \bibinfo {author}
  {\bibfnamefont {N.}~\bibnamefont {Li}}, \bibinfo {author} {\bibfnamefont
  {S.}~\bibnamefont {Wang}}, \bibinfo {author} {\bibfnamefont {Y.}~\bibnamefont
  {Zhao}}, \bibinfo {author} {\bibfnamefont {J.}~\bibnamefont {Tang}}, \bibinfo
  {author} {\bibfnamefont {J.}~\bibnamefont {Liu}}, \bibinfo {author}
  {\bibfnamefont {J.}~\bibnamefont {Tian}}, \bibinfo {author} {\bibfnamefont
  {K.}~\bibnamefont {Watanabe}}, \emph {et~al.},\ }\bibfield  {title} {\bibinfo
  {title} {Correlated states in twisted double bilayer graphene},\ }\href@noop
  {} {\bibfield  {journal} {\bibinfo  {journal} {Nature Physics}\ }\textbf
  {\bibinfo {volume} {16}},\ \bibinfo {pages} {520} (\bibinfo {year}
  {2020})}\BibitemShut {NoStop}%
\bibitem [{\citenamefont {Cao}\ \emph {et~al.}(2020{\natexlab{c}})\citenamefont
  {Cao}, \citenamefont {Rodan-Legrain}, \citenamefont {Rubies-Bigorda},
  \citenamefont {Park}, \citenamefont {Watanabe}, \citenamefont {Taniguchi},\
  and\ \citenamefont {Jarillo-Herrero}}]{cao2019electric}%
  \BibitemOpen
  \bibfield  {author} {\bibinfo {author} {\bibfnamefont {Y.}~\bibnamefont
  {Cao}}, \bibinfo {author} {\bibfnamefont {D.}~\bibnamefont {Rodan-Legrain}},
  \bibinfo {author} {\bibfnamefont {O.}~\bibnamefont {Rubies-Bigorda}},
  \bibinfo {author} {\bibfnamefont {J.~M.}\ \bibnamefont {Park}}, \bibinfo
  {author} {\bibfnamefont {K.}~\bibnamefont {Watanabe}}, \bibinfo {author}
  {\bibfnamefont {T.}~\bibnamefont {Taniguchi}},\ and\ \bibinfo {author}
  {\bibfnamefont {P.}~\bibnamefont {Jarillo-Herrero}},\ }\bibfield  {title}
  {\bibinfo {title} {Author correction: Tunable correlated states and
  spin-polarized phases in twisted bilayer--bilayer graphene},\ }\href
  {https://doi.org/10.1038/s41586-020-2393-7} {\bibfield  {journal} {\bibinfo
  {journal} {Nature}\ }\textbf {\bibinfo {volume} {583}},\ \bibinfo {pages}
  {E27} (\bibinfo {year} {2020}{\natexlab{c}})}\BibitemShut {NoStop}%
\bibitem [{\citenamefont {Burg}\ \emph
  {et~al.}(2019{\natexlab{b}})\citenamefont {Burg}, \citenamefont {Zhu},
  \citenamefont {Taniguchi}, \citenamefont {Watanabe}, \citenamefont
  {MacDonald},\ and\ \citenamefont {Tutuc}}]{tutuc2019}%
  \BibitemOpen
  \bibfield  {author} {\bibinfo {author} {\bibfnamefont {G.~W.}\ \bibnamefont
  {Burg}}, \bibinfo {author} {\bibfnamefont {J.}~\bibnamefont {Zhu}}, \bibinfo
  {author} {\bibfnamefont {T.}~\bibnamefont {Taniguchi}}, \bibinfo {author}
  {\bibfnamefont {K.}~\bibnamefont {Watanabe}}, \bibinfo {author}
  {\bibfnamefont {A.~H.}\ \bibnamefont {MacDonald}},\ and\ \bibinfo {author}
  {\bibfnamefont {E.}~\bibnamefont {Tutuc}},\ }\bibfield  {title} {\bibinfo
  {title} {Correlated insulating states in twisted double bilayer graphene},\
  }\href {https://doi.org/10.1103/PhysRevLett.123.197702} {\bibfield  {journal}
  {\bibinfo  {journal} {Phys. Rev. Lett.}\ }\textbf {\bibinfo {volume} {123}},\
  \bibinfo {pages} {197702} (\bibinfo {year} {2019}{\natexlab{b}})}\BibitemShut
  {NoStop}%
\bibitem [{\citenamefont {Rubio-Verd{\'u}}\ \emph {et~al.}(2022)\citenamefont
  {Rubio-Verd{\'u}}, \citenamefont {Turkel}, \citenamefont {Song},
  \citenamefont {Klebl}, \citenamefont {Samajdar}, \citenamefont {Scheurer},
  \citenamefont {Venderbos}, \citenamefont {Watanabe}, \citenamefont
  {Taniguchi}, \citenamefont {Ochoa} \emph {et~al.}}]{rubioverdu2020universal}%
  \BibitemOpen
  \bibfield  {author} {\bibinfo {author} {\bibfnamefont {C.}~\bibnamefont
  {Rubio-Verd{\'u}}}, \bibinfo {author} {\bibfnamefont {S.}~\bibnamefont
  {Turkel}}, \bibinfo {author} {\bibfnamefont {Y.}~\bibnamefont {Song}},
  \bibinfo {author} {\bibfnamefont {L.}~\bibnamefont {Klebl}}, \bibinfo
  {author} {\bibfnamefont {R.}~\bibnamefont {Samajdar}}, \bibinfo {author}
  {\bibfnamefont {M.~S.}\ \bibnamefont {Scheurer}}, \bibinfo {author}
  {\bibfnamefont {J.~W.}\ \bibnamefont {Venderbos}}, \bibinfo {author}
  {\bibfnamefont {K.}~\bibnamefont {Watanabe}}, \bibinfo {author}
  {\bibfnamefont {T.}~\bibnamefont {Taniguchi}}, \bibinfo {author}
  {\bibfnamefont {H.}~\bibnamefont {Ochoa}}, \emph {et~al.},\ }\bibfield
  {title} {\bibinfo {title} {Moir{\'e} nematic phase in twisted double bilayer
  graphene},\ }\href@noop {} {\bibfield  {journal} {\bibinfo  {journal} {Nature
  Physics}\ }\textbf {\bibinfo {volume} {18}},\ \bibinfo {pages} {196}
  (\bibinfo {year} {2022})}\BibitemShut {NoStop}%
\bibitem [{\citenamefont {Chen}\ \emph
  {et~al.}(2019{\natexlab{a}})\citenamefont {Chen}, \citenamefont {Sharpe},
  \citenamefont {Gallagher}, \citenamefont {Rosen}, \citenamefont {Fox},
  \citenamefont {Jiang}, \citenamefont {Lyu}, \citenamefont {Li}, \citenamefont
  {Watanabe}, \citenamefont {Taniguchi}, \citenamefont {Jung}, \citenamefont
  {Shi}, \citenamefont {Goldhaber-Gordon}, \citenamefont {Zhang},\ and\
  \citenamefont {Wang}}]{Chen2019ABC}%
  \BibitemOpen
  \bibfield  {author} {\bibinfo {author} {\bibfnamefont {G.}~\bibnamefont
  {Chen}}, \bibinfo {author} {\bibfnamefont {A.~L.}\ \bibnamefont {Sharpe}},
  \bibinfo {author} {\bibfnamefont {P.}~\bibnamefont {Gallagher}}, \bibinfo
  {author} {\bibfnamefont {I.~T.}\ \bibnamefont {Rosen}}, \bibinfo {author}
  {\bibfnamefont {E.~J.}\ \bibnamefont {Fox}}, \bibinfo {author} {\bibfnamefont
  {L.}~\bibnamefont {Jiang}}, \bibinfo {author} {\bibfnamefont
  {B.}~\bibnamefont {Lyu}}, \bibinfo {author} {\bibfnamefont {H.}~\bibnamefont
  {Li}}, \bibinfo {author} {\bibfnamefont {K.}~\bibnamefont {Watanabe}},
  \bibinfo {author} {\bibfnamefont {T.}~\bibnamefont {Taniguchi}}, \bibinfo
  {author} {\bibfnamefont {J.}~\bibnamefont {Jung}}, \bibinfo {author}
  {\bibfnamefont {Z.}~\bibnamefont {Shi}}, \bibinfo {author} {\bibfnamefont
  {D.}~\bibnamefont {Goldhaber-Gordon}}, \bibinfo {author} {\bibfnamefont
  {Y.}~\bibnamefont {Zhang}},\ and\ \bibinfo {author} {\bibfnamefont
  {F.}~\bibnamefont {Wang}},\ }\bibfield  {title} {\bibinfo {title} {Signatures
  of tunable superconductivity in a trilayer graphene moir\'e superlattice},\
  }\href@noop {} {\bibfield  {journal} {\bibinfo  {journal} {Nature}\ }\textbf
  {\bibinfo {volume} {572}},\ \bibinfo {pages} {215} (\bibinfo {year}
  {2019}{\natexlab{a}})}\BibitemShut {NoStop}%
\bibitem [{\citenamefont {Chen}\ \emph
  {et~al.}(2019{\natexlab{b}})\citenamefont {Chen}, \citenamefont {Jiang},
  \citenamefont {Wu}, \citenamefont {Lyu}, \citenamefont {Li}, \citenamefont
  {Chittari}, \citenamefont {Watanabe}, \citenamefont {Taniguchi},
  \citenamefont {Shi}, \citenamefont {Jung} \emph {et~al.}}]{chen2019evidence}%
  \BibitemOpen
  \bibfield  {author} {\bibinfo {author} {\bibfnamefont {G.}~\bibnamefont
  {Chen}}, \bibinfo {author} {\bibfnamefont {L.}~\bibnamefont {Jiang}},
  \bibinfo {author} {\bibfnamefont {S.}~\bibnamefont {Wu}}, \bibinfo {author}
  {\bibfnamefont {B.}~\bibnamefont {Lyu}}, \bibinfo {author} {\bibfnamefont
  {H.}~\bibnamefont {Li}}, \bibinfo {author} {\bibfnamefont {B.~L.}\
  \bibnamefont {Chittari}}, \bibinfo {author} {\bibfnamefont {K.}~\bibnamefont
  {Watanabe}}, \bibinfo {author} {\bibfnamefont {T.}~\bibnamefont {Taniguchi}},
  \bibinfo {author} {\bibfnamefont {Z.}~\bibnamefont {Shi}}, \bibinfo {author}
  {\bibfnamefont {J.}~\bibnamefont {Jung}}, \emph {et~al.},\ }\bibfield
  {title} {\bibinfo {title} {Evidence of a gate-tunable mott insulator in a
  trilayer graphene moir{\'e} superlattice},\ }\href@noop {} {\bibfield
  {journal} {\bibinfo  {journal} {Nature Physics}\ }\textbf {\bibinfo {volume}
  {15}},\ \bibinfo {pages} {237} (\bibinfo {year}
  {2019}{\natexlab{b}})}\BibitemShut {NoStop}%
\bibitem [{\citenamefont {Chen}\ \emph {et~al.}(2020)\citenamefont {Chen},
  \citenamefont {Sharpe}, \citenamefont {Fox}, \citenamefont {Zhang},
  \citenamefont {Wang}, \citenamefont {Jiang}, \citenamefont {Lyu},
  \citenamefont {Li}, \citenamefont {Watanabe}, \citenamefont {Taniguchi} \emph
  {et~al.}}]{chen2020tunable}%
  \BibitemOpen
  \bibfield  {author} {\bibinfo {author} {\bibfnamefont {G.}~\bibnamefont
  {Chen}}, \bibinfo {author} {\bibfnamefont {A.~L.}\ \bibnamefont {Sharpe}},
  \bibinfo {author} {\bibfnamefont {E.~J.}\ \bibnamefont {Fox}}, \bibinfo
  {author} {\bibfnamefont {Y.-H.}\ \bibnamefont {Zhang}}, \bibinfo {author}
  {\bibfnamefont {S.}~\bibnamefont {Wang}}, \bibinfo {author} {\bibfnamefont
  {L.}~\bibnamefont {Jiang}}, \bibinfo {author} {\bibfnamefont
  {B.}~\bibnamefont {Lyu}}, \bibinfo {author} {\bibfnamefont {H.}~\bibnamefont
  {Li}}, \bibinfo {author} {\bibfnamefont {K.}~\bibnamefont {Watanabe}},
  \bibinfo {author} {\bibfnamefont {T.}~\bibnamefont {Taniguchi}}, \emph
  {et~al.},\ }\bibfield  {title} {\bibinfo {title} {Tunable correlated chern
  insulator and ferromagnetism in a moir{\'e} superlattice},\ }\href@noop {}
  {\bibfield  {journal} {\bibinfo  {journal} {Nature}\ }\textbf {\bibinfo
  {volume} {579}},\ \bibinfo {pages} {56} (\bibinfo {year} {2020})}\BibitemShut
  {NoStop}%
\bibitem [{\citenamefont {Park}\ \emph
  {et~al.}(2021{\natexlab{b}})\citenamefont {Park}, \citenamefont {Cao},
  \citenamefont {Xia}, \citenamefont {Sun}, \citenamefont {Watanabe},
  \citenamefont {Taniguchi},\ and\ \citenamefont
  {Jarillo-Herrero}}]{park2021multi}%
  \BibitemOpen
  \bibfield  {author} {\bibinfo {author} {\bibfnamefont {J.~M.}\ \bibnamefont
  {Park}}, \bibinfo {author} {\bibfnamefont {Y.}~\bibnamefont {Cao}}, \bibinfo
  {author} {\bibfnamefont {L.}~\bibnamefont {Xia}}, \bibinfo {author}
  {\bibfnamefont {S.}~\bibnamefont {Sun}}, \bibinfo {author} {\bibfnamefont
  {K.}~\bibnamefont {Watanabe}}, \bibinfo {author} {\bibfnamefont
  {T.}~\bibnamefont {Taniguchi}},\ and\ \bibinfo {author} {\bibfnamefont
  {P.}~\bibnamefont {Jarillo-Herrero}},\ }\href
  {https://doi.org/10.48550/ARXIV.2112.10760} {\bibinfo {title} {Magic-angle
  multilayer graphene: A robust family of moiré superconductors}} (\bibinfo
  {year} {2021}{\natexlab{b}})\BibitemShut {NoStop}%
\bibitem [{\citenamefont {Zhang}\ \emph {et~al.}(2021)\citenamefont {Zhang},
  \citenamefont {Polski}, \citenamefont {Lewandowski}, \citenamefont {Thomson},
  \citenamefont {Peng}, \citenamefont {Choi}, \citenamefont {Kim},
  \citenamefont {Watanabe}, \citenamefont {Taniguchi}, \citenamefont {Alicea},
  \citenamefont {von Oppen}, \citenamefont {Refael},\ and\ \citenamefont
  {Nadj-Perge}}]{zhang2021multi}%
  \BibitemOpen
  \bibfield  {author} {\bibinfo {author} {\bibfnamefont {Y.}~\bibnamefont
  {Zhang}}, \bibinfo {author} {\bibfnamefont {R.}~\bibnamefont {Polski}},
  \bibinfo {author} {\bibfnamefont {C.}~\bibnamefont {Lewandowski}}, \bibinfo
  {author} {\bibfnamefont {A.}~\bibnamefont {Thomson}}, \bibinfo {author}
  {\bibfnamefont {Y.}~\bibnamefont {Peng}}, \bibinfo {author} {\bibfnamefont
  {Y.}~\bibnamefont {Choi}}, \bibinfo {author} {\bibfnamefont {H.}~\bibnamefont
  {Kim}}, \bibinfo {author} {\bibfnamefont {K.}~\bibnamefont {Watanabe}},
  \bibinfo {author} {\bibfnamefont {T.}~\bibnamefont {Taniguchi}}, \bibinfo
  {author} {\bibfnamefont {J.}~\bibnamefont {Alicea}}, \bibinfo {author}
  {\bibfnamefont {F.}~\bibnamefont {von Oppen}}, \bibinfo {author}
  {\bibfnamefont {G.}~\bibnamefont {Refael}},\ and\ \bibinfo {author}
  {\bibfnamefont {S.}~\bibnamefont {Nadj-Perge}},\ }\href
  {https://doi.org/10.48550/ARXIV.2112.09270} {\bibinfo {title} {Ascendance of
  superconductivity in magic-angle graphene multilayers}} (\bibinfo {year}
  {2021})\BibitemShut {NoStop}%
\bibitem [{\citenamefont {Burg}\ \emph {et~al.}(2022)\citenamefont {Burg},
  \citenamefont {Khalaf}, \citenamefont {Wang}, \citenamefont {Watanabe},
  \citenamefont {Taniguchi},\ and\ \citenamefont {Tutuc}}]{burg2021multi}%
  \BibitemOpen
  \bibfield  {author} {\bibinfo {author} {\bibfnamefont {G.~W.}\ \bibnamefont
  {Burg}}, \bibinfo {author} {\bibfnamefont {E.}~\bibnamefont {Khalaf}},
  \bibinfo {author} {\bibfnamefont {Y.}~\bibnamefont {Wang}}, \bibinfo {author}
  {\bibfnamefont {K.}~\bibnamefont {Watanabe}}, \bibinfo {author}
  {\bibfnamefont {T.}~\bibnamefont {Taniguchi}},\ and\ \bibinfo {author}
  {\bibfnamefont {E.}~\bibnamefont {Tutuc}},\ }\href
  {https://doi.org/10.48550/ARXIV.2201.01637} {\bibinfo {title} {Emergence of
  correlations at the edge of the magic angle regime in alternating twist
  quadrilayer graphene}} (\bibinfo {year} {2022})\BibitemShut {NoStop}%
\bibitem [{\citenamefont {Kerelsky}\ \emph {et~al.}(2021)\citenamefont
  {Kerelsky}, \citenamefont {Rubio-Verdú}, \citenamefont {Xian}, \citenamefont
  {Kennes}, \citenamefont {Halbertal}, \citenamefont {Finney}, \citenamefont
  {Song}, \citenamefont {Turkel}, \citenamefont {Wang}, \citenamefont
  {Watanabe}, \citenamefont {Taniguchi}, \citenamefont {Hone}, \citenamefont
  {Dean}, \citenamefont {Basov}, \citenamefont {Rubio},\ and\ \citenamefont
  {Pasupathy}}]{kerelsky2021moireless}%
  \BibitemOpen
  \bibfield  {author} {\bibinfo {author} {\bibfnamefont {A.}~\bibnamefont
  {Kerelsky}}, \bibinfo {author} {\bibfnamefont {C.}~\bibnamefont
  {Rubio-Verdú}}, \bibinfo {author} {\bibfnamefont {L.}~\bibnamefont {Xian}},
  \bibinfo {author} {\bibfnamefont {D.~M.}\ \bibnamefont {Kennes}}, \bibinfo
  {author} {\bibfnamefont {D.}~\bibnamefont {Halbertal}}, \bibinfo {author}
  {\bibfnamefont {N.}~\bibnamefont {Finney}}, \bibinfo {author} {\bibfnamefont
  {L.}~\bibnamefont {Song}}, \bibinfo {author} {\bibfnamefont {S.}~\bibnamefont
  {Turkel}}, \bibinfo {author} {\bibfnamefont {L.}~\bibnamefont {Wang}},
  \bibinfo {author} {\bibfnamefont {K.}~\bibnamefont {Watanabe}}, \bibinfo
  {author} {\bibfnamefont {T.}~\bibnamefont {Taniguchi}}, \bibinfo {author}
  {\bibfnamefont {J.}~\bibnamefont {Hone}}, \bibinfo {author} {\bibfnamefont
  {C.}~\bibnamefont {Dean}}, \bibinfo {author} {\bibfnamefont {D.~N.}\
  \bibnamefont {Basov}}, \bibinfo {author} {\bibfnamefont {A.}~\bibnamefont
  {Rubio}},\ and\ \bibinfo {author} {\bibfnamefont {A.~N.}\ \bibnamefont
  {Pasupathy}},\ }\bibfield  {title} {\bibinfo {title} {Moir\'eless
  correlations in abca graphene},\ }\href
  {https://doi.org/10.1073/pnas.2017366118} {\bibfield  {journal} {\bibinfo
  {journal} {Proceedings of the National Academy of Sciences}\ }\textbf
  {\bibinfo {volume} {118}},\ \bibinfo {pages} {e2017366118} (\bibinfo {year}
  {2021})},\ \Eprint
  {https://arxiv.org/abs/https://www.pnas.org/doi/pdf/10.1073/pnas.2017366118}
  {https://www.pnas.org/doi/pdf/10.1073/pnas.2017366118} \BibitemShut {NoStop}%
\bibitem [{\citenamefont {Liu}\ \emph {et~al.}(2022)\citenamefont {Liu},
  \citenamefont {Zhang}, \citenamefont {Watanabe}, \citenamefont {Taniguchi},\
  and\ \citenamefont {Li}}]{liu2022isospin}%
  \BibitemOpen
  \bibfield  {author} {\bibinfo {author} {\bibfnamefont {X.}~\bibnamefont
  {Liu}}, \bibinfo {author} {\bibfnamefont {N.~J.}\ \bibnamefont {Zhang}},
  \bibinfo {author} {\bibfnamefont {K.}~\bibnamefont {Watanabe}}, \bibinfo
  {author} {\bibfnamefont {T.}~\bibnamefont {Taniguchi}},\ and\ \bibinfo
  {author} {\bibfnamefont {J.}~\bibnamefont {Li}},\ }\bibfield  {title}
  {\bibinfo {title} {Isospin order in superconducting magic-angle twisted
  trilayer graphene},\ }\href@noop {} {\bibfield  {journal} {\bibinfo
  {journal} {Nature Physics}\ ,\ \bibinfo {pages} {1}} (\bibinfo {year}
  {2022})}\BibitemShut {NoStop}%
\bibitem [{\citenamefont {Wang}\ \emph {et~al.}(2020)\citenamefont {Wang},
  \citenamefont {Shih}, \citenamefont {Ghiotto}, \citenamefont {Xian},
  \citenamefont {Rhodes}, \citenamefont {Tan}, \citenamefont {Claassen},
  \citenamefont {Kennes}, \citenamefont {Bai}, \citenamefont {Kim} \emph
  {et~al.}}]{wang2020correlated}%
  \BibitemOpen
  \bibfield  {author} {\bibinfo {author} {\bibfnamefont {L.}~\bibnamefont
  {Wang}}, \bibinfo {author} {\bibfnamefont {E.-M.}\ \bibnamefont {Shih}},
  \bibinfo {author} {\bibfnamefont {A.}~\bibnamefont {Ghiotto}}, \bibinfo
  {author} {\bibfnamefont {L.}~\bibnamefont {Xian}}, \bibinfo {author}
  {\bibfnamefont {D.~A.}\ \bibnamefont {Rhodes}}, \bibinfo {author}
  {\bibfnamefont {C.}~\bibnamefont {Tan}}, \bibinfo {author} {\bibfnamefont
  {M.}~\bibnamefont {Claassen}}, \bibinfo {author} {\bibfnamefont {D.~M.}\
  \bibnamefont {Kennes}}, \bibinfo {author} {\bibfnamefont {Y.}~\bibnamefont
  {Bai}}, \bibinfo {author} {\bibfnamefont {B.}~\bibnamefont {Kim}}, \emph
  {et~al.},\ }\bibfield  {title} {\bibinfo {title} {Correlated electronic
  phases in twisted bilayer transition metal dichalcogenides},\ }\href@noop {}
  {\bibfield  {journal} {\bibinfo  {journal} {Nature materials}\ }\textbf
  {\bibinfo {volume} {19}},\ \bibinfo {pages} {861} (\bibinfo {year}
  {2020})}\BibitemShut {NoStop}%
\bibitem [{\citenamefont {Ghiotto}\ \emph {et~al.}(2021)\citenamefont
  {Ghiotto}, \citenamefont {Shih}, \citenamefont {Pereira}, \citenamefont
  {Rhodes}, \citenamefont {Kim}, \citenamefont {Zang}, \citenamefont {Millis},
  \citenamefont {Watanabe}, \citenamefont {Taniguchi}, \citenamefont {Hone}
  \emph {et~al.}}]{ghiotto2021quantum}%
  \BibitemOpen
  \bibfield  {author} {\bibinfo {author} {\bibfnamefont {A.}~\bibnamefont
  {Ghiotto}}, \bibinfo {author} {\bibfnamefont {E.-M.}\ \bibnamefont {Shih}},
  \bibinfo {author} {\bibfnamefont {G.~S.}\ \bibnamefont {Pereira}}, \bibinfo
  {author} {\bibfnamefont {D.~A.}\ \bibnamefont {Rhodes}}, \bibinfo {author}
  {\bibfnamefont {B.}~\bibnamefont {Kim}}, \bibinfo {author} {\bibfnamefont
  {J.}~\bibnamefont {Zang}}, \bibinfo {author} {\bibfnamefont {A.~J.}\
  \bibnamefont {Millis}}, \bibinfo {author} {\bibfnamefont {K.}~\bibnamefont
  {Watanabe}}, \bibinfo {author} {\bibfnamefont {T.}~\bibnamefont {Taniguchi}},
  \bibinfo {author} {\bibfnamefont {J.~C.}\ \bibnamefont {Hone}}, \emph
  {et~al.},\ }\bibfield  {title} {\bibinfo {title} {Quantum criticality in
  twisted transition metal dichalcogenides},\ }\href@noop {} {\bibfield
  {journal} {\bibinfo  {journal} {Nature}\ }\textbf {\bibinfo {volume} {597}},\
  \bibinfo {pages} {345} (\bibinfo {year} {2021})}\BibitemShut {NoStop}%
\bibitem [{\citenamefont {Tang}\ \emph {et~al.}(2020)\citenamefont {Tang},
  \citenamefont {Li}, \citenamefont {Li}, \citenamefont {Xu}, \citenamefont
  {Liu}, \citenamefont {Barmak}, \citenamefont {Watanabe}, \citenamefont
  {Taniguchi}, \citenamefont {MacDonald}, \citenamefont {Shan},\ and\
  \citenamefont {Mak}}]{tang2020}%
  \BibitemOpen
  \bibfield  {author} {\bibinfo {author} {\bibfnamefont {Y.}~\bibnamefont
  {Tang}}, \bibinfo {author} {\bibfnamefont {L.}~\bibnamefont {Li}}, \bibinfo
  {author} {\bibfnamefont {T.}~\bibnamefont {Li}}, \bibinfo {author}
  {\bibfnamefont {Y.}~\bibnamefont {Xu}}, \bibinfo {author} {\bibfnamefont
  {S.}~\bibnamefont {Liu}}, \bibinfo {author} {\bibfnamefont {K.}~\bibnamefont
  {Barmak}}, \bibinfo {author} {\bibfnamefont {K.}~\bibnamefont {Watanabe}},
  \bibinfo {author} {\bibfnamefont {T.}~\bibnamefont {Taniguchi}}, \bibinfo
  {author} {\bibfnamefont {A.~H.}\ \bibnamefont {MacDonald}}, \bibinfo {author}
  {\bibfnamefont {J.}~\bibnamefont {Shan}},\ and\ \bibinfo {author}
  {\bibfnamefont {K.~F.}\ \bibnamefont {Mak}},\ }\bibfield  {title} {\bibinfo
  {title} {Simulation of hubbard model physics in wse$_2$/ws$_2$ moir\'e
  superlattices},\ }\href@noop {} {\bibfield  {journal} {\bibinfo  {journal}
  {Nature}\ }\textbf {\bibinfo {volume} {579}},\ \bibinfo {pages} {353}
  (\bibinfo {year} {2020})}\BibitemShut {NoStop}%
\bibitem [{\citenamefont {Jin}\ \emph {et~al.}(2021)\citenamefont {Jin},
  \citenamefont {Tao}, \citenamefont {Li}, \citenamefont {Xu}, \citenamefont
  {Tang}, \citenamefont {Zhu}, \citenamefont {Liu}, \citenamefont {Watanabe},
  \citenamefont {Taniguchi}, \citenamefont {Hone} \emph
  {et~al.}}]{jin2021stripe}%
  \BibitemOpen
  \bibfield  {author} {\bibinfo {author} {\bibfnamefont {C.}~\bibnamefont
  {Jin}}, \bibinfo {author} {\bibfnamefont {Z.}~\bibnamefont {Tao}}, \bibinfo
  {author} {\bibfnamefont {T.}~\bibnamefont {Li}}, \bibinfo {author}
  {\bibfnamefont {Y.}~\bibnamefont {Xu}}, \bibinfo {author} {\bibfnamefont
  {Y.}~\bibnamefont {Tang}}, \bibinfo {author} {\bibfnamefont {J.}~\bibnamefont
  {Zhu}}, \bibinfo {author} {\bibfnamefont {S.}~\bibnamefont {Liu}}, \bibinfo
  {author} {\bibfnamefont {K.}~\bibnamefont {Watanabe}}, \bibinfo {author}
  {\bibfnamefont {T.}~\bibnamefont {Taniguchi}}, \bibinfo {author}
  {\bibfnamefont {J.~C.}\ \bibnamefont {Hone}}, \emph {et~al.},\ }\bibfield
  {title} {\bibinfo {title} {Stripe phases in wse2/ws2 moir{\'e}
  superlattices},\ }\href@noop {} {\bibfield  {journal} {\bibinfo  {journal}
  {Nature Materials}\ }\textbf {\bibinfo {volume} {20}},\ \bibinfo {pages}
  {940} (\bibinfo {year} {2021})}\BibitemShut {NoStop}%
\bibitem [{\citenamefont {Jin}\ \emph {et~al.}(2019)\citenamefont {Jin},
  \citenamefont {Regan}, \citenamefont {Yan}, \citenamefont {Iqbal
  Bakti~Utama}, \citenamefont {Wang}, \citenamefont {Zhao}, \citenamefont
  {Qin}, \citenamefont {Yang}, \citenamefont {Zheng}, \citenamefont {Shi} \emph
  {et~al.}}]{jin2019observation}%
  \BibitemOpen
  \bibfield  {author} {\bibinfo {author} {\bibfnamefont {C.}~\bibnamefont
  {Jin}}, \bibinfo {author} {\bibfnamefont {E.~C.}\ \bibnamefont {Regan}},
  \bibinfo {author} {\bibfnamefont {A.}~\bibnamefont {Yan}}, \bibinfo {author}
  {\bibfnamefont {M.}~\bibnamefont {Iqbal Bakti~Utama}}, \bibinfo {author}
  {\bibfnamefont {D.}~\bibnamefont {Wang}}, \bibinfo {author} {\bibfnamefont
  {S.}~\bibnamefont {Zhao}}, \bibinfo {author} {\bibfnamefont {Y.}~\bibnamefont
  {Qin}}, \bibinfo {author} {\bibfnamefont {S.}~\bibnamefont {Yang}}, \bibinfo
  {author} {\bibfnamefont {Z.}~\bibnamefont {Zheng}}, \bibinfo {author}
  {\bibfnamefont {S.}~\bibnamefont {Shi}}, \emph {et~al.},\ }\bibfield  {title}
  {\bibinfo {title} {Observation of moir{\'e} excitons in wse2/ws2
  heterostructure superlattices},\ }\href@noop {} {\bibfield  {journal}
  {\bibinfo  {journal} {Nature}\ }\textbf {\bibinfo {volume} {567}},\ \bibinfo
  {pages} {76} (\bibinfo {year} {2019})}\BibitemShut {NoStop}%
\bibitem [{\citenamefont {{Wang}}\ \emph {et~al.}(2019)\citenamefont {{Wang}},
  \citenamefont {{Rhodes}}, \citenamefont {{Watanabe}}, \citenamefont
  {{Taniguchi}}, \citenamefont {{Hone}}, \citenamefont {{Shan}},\ and\
  \citenamefont {{Mak}}}]{2019Natur.574...76W}%
  \BibitemOpen
  \bibfield  {author} {\bibinfo {author} {\bibfnamefont {Z.}~\bibnamefont
  {{Wang}}}, \bibinfo {author} {\bibfnamefont {D.~A.}\ \bibnamefont
  {{Rhodes}}}, \bibinfo {author} {\bibfnamefont {K.}~\bibnamefont
  {{Watanabe}}}, \bibinfo {author} {\bibfnamefont {T.}~\bibnamefont
  {{Taniguchi}}}, \bibinfo {author} {\bibfnamefont {J.~C.}\ \bibnamefont
  {{Hone}}}, \bibinfo {author} {\bibfnamefont {J.}~\bibnamefont {{Shan}}},\
  and\ \bibinfo {author} {\bibfnamefont {K.~F.}\ \bibnamefont {{Mak}}},\
  }\bibfield  {title} {\bibinfo {title} {{Evidence of high-temperature exciton
  condensation in two-dimensional atomic double layers}},\ }\href
  {https://doi.org/10.1038/s41586-019-1591-7} {\bibfield  {journal} {\bibinfo
  {journal} {\nat}\ }\textbf {\bibinfo {volume} {574}},\ \bibinfo {pages} {76}
  (\bibinfo {year} {2019})},\ \Eprint {https://arxiv.org/abs/2103.16407}
  {arXiv:2103.16407 [cond-mat.mes-hall]} \BibitemShut {NoStop}%
\bibitem [{\citenamefont {Shimazaki}\ \emph {et~al.}(2020)\citenamefont
  {Shimazaki}, \citenamefont {Schwartz}, \citenamefont {Watanabe},
  \citenamefont {Taniguchi}, \citenamefont {Kroner},\ and\ \citenamefont
  {Imamo{\u{g}}lu}}]{shimazaki2020strongly}%
  \BibitemOpen
  \bibfield  {author} {\bibinfo {author} {\bibfnamefont {Y.}~\bibnamefont
  {Shimazaki}}, \bibinfo {author} {\bibfnamefont {I.}~\bibnamefont {Schwartz}},
  \bibinfo {author} {\bibfnamefont {K.}~\bibnamefont {Watanabe}}, \bibinfo
  {author} {\bibfnamefont {T.}~\bibnamefont {Taniguchi}}, \bibinfo {author}
  {\bibfnamefont {M.}~\bibnamefont {Kroner}},\ and\ \bibinfo {author}
  {\bibfnamefont {A.}~\bibnamefont {Imamo{\u{g}}lu}},\ }\bibfield  {title}
  {\bibinfo {title} {Strongly correlated electrons and hybrid excitons in a
  moir{\'e} heterostructure},\ }\href@noop {} {\bibfield  {journal} {\bibinfo
  {journal} {Nature}\ }\textbf {\bibinfo {volume} {580}},\ \bibinfo {pages}
  {472} (\bibinfo {year} {2020})}\BibitemShut {NoStop}%
\bibitem [{\citenamefont {Regan}\ \emph {et~al.}(2020)\citenamefont {Regan},
  \citenamefont {Wang}, \citenamefont {Jin}, \citenamefont {Bakti~Utama},
  \citenamefont {Gao}, \citenamefont {Wei}, \citenamefont {Zhao}, \citenamefont
  {Zhao}, \citenamefont {Zhang}, \citenamefont {Yumigeta} \emph
  {et~al.}}]{regan2020mott}%
  \BibitemOpen
  \bibfield  {author} {\bibinfo {author} {\bibfnamefont {E.~C.}\ \bibnamefont
  {Regan}}, \bibinfo {author} {\bibfnamefont {D.}~\bibnamefont {Wang}},
  \bibinfo {author} {\bibfnamefont {C.}~\bibnamefont {Jin}}, \bibinfo {author}
  {\bibfnamefont {M.~I.}\ \bibnamefont {Bakti~Utama}}, \bibinfo {author}
  {\bibfnamefont {B.}~\bibnamefont {Gao}}, \bibinfo {author} {\bibfnamefont
  {X.}~\bibnamefont {Wei}}, \bibinfo {author} {\bibfnamefont {S.}~\bibnamefont
  {Zhao}}, \bibinfo {author} {\bibfnamefont {W.}~\bibnamefont {Zhao}}, \bibinfo
  {author} {\bibfnamefont {Z.}~\bibnamefont {Zhang}}, \bibinfo {author}
  {\bibfnamefont {K.}~\bibnamefont {Yumigeta}}, \emph {et~al.},\ }\bibfield
  {title} {\bibinfo {title} {Mott and generalized wigner crystal states in
  wse2/ws2 moir{\'e} superlattices},\ }\href@noop {} {\bibfield  {journal}
  {\bibinfo  {journal} {Nature}\ }\textbf {\bibinfo {volume} {579}},\ \bibinfo
  {pages} {359} (\bibinfo {year} {2020})}\BibitemShut {NoStop}%
\bibitem [{\citenamefont {Li}\ \emph {et~al.}(2021)\citenamefont {Li},
  \citenamefont {Jiang}, \citenamefont {Shen}, \citenamefont {Zhang},
  \citenamefont {Li}, \citenamefont {Tao}, \citenamefont {Devakul},
  \citenamefont {Watanabe}, \citenamefont {Taniguchi}, \citenamefont {Fu} \emph
  {et~al.}}]{li2021quantum}%
  \BibitemOpen
  \bibfield  {author} {\bibinfo {author} {\bibfnamefont {T.}~\bibnamefont
  {Li}}, \bibinfo {author} {\bibfnamefont {S.}~\bibnamefont {Jiang}}, \bibinfo
  {author} {\bibfnamefont {B.}~\bibnamefont {Shen}}, \bibinfo {author}
  {\bibfnamefont {Y.}~\bibnamefont {Zhang}}, \bibinfo {author} {\bibfnamefont
  {L.}~\bibnamefont {Li}}, \bibinfo {author} {\bibfnamefont {Z.}~\bibnamefont
  {Tao}}, \bibinfo {author} {\bibfnamefont {T.}~\bibnamefont {Devakul}},
  \bibinfo {author} {\bibfnamefont {K.}~\bibnamefont {Watanabe}}, \bibinfo
  {author} {\bibfnamefont {T.}~\bibnamefont {Taniguchi}}, \bibinfo {author}
  {\bibfnamefont {L.}~\bibnamefont {Fu}}, \emph {et~al.},\ }\bibfield  {title}
  {\bibinfo {title} {Quantum anomalous hall effect from intertwined moir{\'e}
  bands},\ }\href@noop {} {\bibfield  {journal} {\bibinfo  {journal} {Nature}\
  }\textbf {\bibinfo {volume} {600}},\ \bibinfo {pages} {641} (\bibinfo {year}
  {2021})}\BibitemShut {NoStop}%
\bibitem [{\citenamefont {Hsu}\ \emph {et~al.}(2021)\citenamefont {Hsu},
  \citenamefont {Wu},\ and\ \citenamefont {Das~Sarma}}]{PhysRevB.104.195134}%
  \BibitemOpen
  \bibfield  {author} {\bibinfo {author} {\bibfnamefont {Y.-T.}\ \bibnamefont
  {Hsu}}, \bibinfo {author} {\bibfnamefont {F.}~\bibnamefont {Wu}},\ and\
  \bibinfo {author} {\bibfnamefont {S.}~\bibnamefont {Das~Sarma}},\ }\bibfield
  {title} {\bibinfo {title} {Spin-valley locked instabilities in moir\'e
  transition metal dichalcogenides with conventional and higher-order van hove
  singularities},\ }\href {https://doi.org/10.1103/PhysRevB.104.195134}
  {\bibfield  {journal} {\bibinfo  {journal} {Phys. Rev. B}\ }\textbf {\bibinfo
  {volume} {104}},\ \bibinfo {pages} {195134} (\bibinfo {year}
  {2021})}\BibitemShut {NoStop}%
\bibitem [{\citenamefont {Wu}\ \emph {et~al.}(2022)\citenamefont {Wu},
  \citenamefont {Wu},\ and\ \citenamefont {Yao}}]{wu2022pair}%
  \BibitemOpen
  \bibfield  {author} {\bibinfo {author} {\bibfnamefont {Y.-M.}\ \bibnamefont
  {Wu}}, \bibinfo {author} {\bibfnamefont {Z.~W.}\ \bibnamefont {Wu}},\ and\
  \bibinfo {author} {\bibfnamefont {H.}~\bibnamefont {Yao}},\ }\href@noop {}
  {\bibinfo {title} {Pair-density-wave and chiral superconductivity in twisted
  bilayer transition-metal-dichalcogenides}} (\bibinfo {year} {2022}),\ \Eprint
  {https://arxiv.org/abs/2203.05480} {arXiv:2203.05480 [cond-mat.supr-con]}
  \BibitemShut {NoStop}%
\bibitem [{\citenamefont {Zang}\ \emph {et~al.}(2021)\citenamefont {Zang},
  \citenamefont {Wang}, \citenamefont {Cano},\ and\ \citenamefont
  {Millis}}]{andy2021hartree}%
  \BibitemOpen
  \bibfield  {author} {\bibinfo {author} {\bibfnamefont {J.}~\bibnamefont
  {Zang}}, \bibinfo {author} {\bibfnamefont {J.}~\bibnamefont {Wang}}, \bibinfo
  {author} {\bibfnamefont {J.}~\bibnamefont {Cano}},\ and\ \bibinfo {author}
  {\bibfnamefont {A.~J.}\ \bibnamefont {Millis}},\ }\bibfield  {title}
  {\bibinfo {title} {Hartree-fock study of the moir\'e hubbard model for
  twisted bilayer transition metal dichalcogenides},\ }\href
  {https://doi.org/10.1103/PhysRevB.104.075150} {\bibfield  {journal} {\bibinfo
   {journal} {Phys. Rev. B}\ }\textbf {\bibinfo {volume} {104}},\ \bibinfo
  {pages} {075150} (\bibinfo {year} {2021})}\BibitemShut {NoStop}%
\bibitem [{\citenamefont {Wu}\ \emph {et~al.}(2019)\citenamefont {Wu},
  \citenamefont {Lovorn}, \citenamefont {Tutuc}, \citenamefont {Martin},\ and\
  \citenamefont {MacDonald}}]{PhysRevLett.122.086402}%
  \BibitemOpen
  \bibfield  {author} {\bibinfo {author} {\bibfnamefont {F.}~\bibnamefont
  {Wu}}, \bibinfo {author} {\bibfnamefont {T.}~\bibnamefont {Lovorn}}, \bibinfo
  {author} {\bibfnamefont {E.}~\bibnamefont {Tutuc}}, \bibinfo {author}
  {\bibfnamefont {I.}~\bibnamefont {Martin}},\ and\ \bibinfo {author}
  {\bibfnamefont {A.~H.}\ \bibnamefont {MacDonald}},\ }\bibfield  {title}
  {\bibinfo {title} {Topological insulators in twisted transition metal
  dichalcogenide homobilayers},\ }\href
  {https://doi.org/10.1103/PhysRevLett.122.086402} {\bibfield  {journal}
  {\bibinfo  {journal} {Phys. Rev. Lett.}\ }\textbf {\bibinfo {volume} {122}},\
  \bibinfo {pages} {086402} (\bibinfo {year} {2019})}\BibitemShut {NoStop}%
\bibitem [{\citenamefont {Pan}\ \emph {et~al.}(2020)\citenamefont {Pan},
  \citenamefont {Wu},\ and\ \citenamefont
  {Das~Sarma}}]{PhysRevResearch.2.033087}%
  \BibitemOpen
  \bibfield  {author} {\bibinfo {author} {\bibfnamefont {H.}~\bibnamefont
  {Pan}}, \bibinfo {author} {\bibfnamefont {F.}~\bibnamefont {Wu}},\ and\
  \bibinfo {author} {\bibfnamefont {S.}~\bibnamefont {Das~Sarma}},\ }\bibfield
  {title} {\bibinfo {title} {Band topology, hubbard model, heisenberg model,
  and dzyaloshinskii-moriya interaction in twisted bilayer
  ${\mathrm{wse}}_{2}$},\ }\href
  {https://doi.org/10.1103/PhysRevResearch.2.033087} {\bibfield  {journal}
  {\bibinfo  {journal} {Phys. Rev. Research}\ }\textbf {\bibinfo {volume}
  {2}},\ \bibinfo {pages} {033087} (\bibinfo {year} {2020})}\BibitemShut
  {NoStop}%
\bibitem [{\citenamefont {Kiese}\ \emph {et~al.}(2022)\citenamefont {Kiese},
  \citenamefont {He}, \citenamefont {Hickey}, \citenamefont {Rubio},\ and\
  \citenamefont {Kennes}}]{doi:10.1063/5.0077901}%
  \BibitemOpen
  \bibfield  {author} {\bibinfo {author} {\bibfnamefont {D.}~\bibnamefont
  {Kiese}}, \bibinfo {author} {\bibfnamefont {Y.}~\bibnamefont {He}}, \bibinfo
  {author} {\bibfnamefont {C.}~\bibnamefont {Hickey}}, \bibinfo {author}
  {\bibfnamefont {A.}~\bibnamefont {Rubio}},\ and\ \bibinfo {author}
  {\bibfnamefont {D.~M.}\ \bibnamefont {Kennes}},\ }\bibfield  {title}
  {\bibinfo {title} {Tmds as a platform for spin liquid physics: A strong
  coupling study of twisted bilayer wse2},\ }\href
  {https://doi.org/10.1063/5.0077901} {\bibfield  {journal} {\bibinfo
  {journal} {APL Materials}\ }\textbf {\bibinfo {volume} {10}},\ \bibinfo
  {pages} {031113} (\bibinfo {year} {2022})},\ \Eprint
  {https://arxiv.org/abs/https://doi.org/10.1063/5.0077901}
  {https://doi.org/10.1063/5.0077901} \BibitemShut {NoStop}%
\bibitem [{\citenamefont {Klebl}\ \emph
  {et~al.}(2022{\natexlab{a}})\citenamefont {Klebl}, \citenamefont {Fischer},
  \citenamefont {Claassen}, \citenamefont {Scherer},\ and\ \citenamefont
  {Kennes}}]{supplement}%
  \BibitemOpen
  \bibfield  {author} {\bibinfo {author} {\bibfnamefont {L.}~\bibnamefont
  {Klebl}}, \bibinfo {author} {\bibfnamefont {A.}~\bibnamefont {Fischer}},
  \bibinfo {author} {\bibfnamefont {L.}~\bibnamefont {Claassen}}, \bibinfo
  {author} {\bibfnamefont {M.~M.}\ \bibnamefont {Scherer}},\ and\ \bibinfo
  {author} {\bibfnamefont {D.~M.}\ \bibnamefont {Kennes}},\ }\href@noop {}
  {\bibinfo {title} {Supplemental material}} (\bibinfo {year}
  {2022}{\natexlab{a}})\BibitemShut {NoStop}%
\bibitem [{\citenamefont {Wu}\ \emph {et~al.}(2018)\citenamefont {Wu},
  \citenamefont {Lovorn}, \citenamefont {Tutuc},\ and\ \citenamefont
  {MacDonald}}]{Wu17}%
  \BibitemOpen
  \bibfield  {author} {\bibinfo {author} {\bibfnamefont {F.}~\bibnamefont
  {Wu}}, \bibinfo {author} {\bibfnamefont {T.}~\bibnamefont {Lovorn}}, \bibinfo
  {author} {\bibfnamefont {E.}~\bibnamefont {Tutuc}},\ and\ \bibinfo {author}
  {\bibfnamefont {A.~H.}\ \bibnamefont {MacDonald}},\ }\bibfield  {title}
  {\bibinfo {title} {Hubbard model physics in transition metal dichalcogenide
  moir\'e bands},\ }\href {https://doi.org/10.1103/PhysRevLett.121.026402}
  {\bibfield  {journal} {\bibinfo  {journal} {Phys. Rev. Lett.}\ }\textbf
  {\bibinfo {volume} {121}},\ \bibinfo {pages} {026402} (\bibinfo {year}
  {2018})}\BibitemShut {NoStop}%
\bibitem [{\citenamefont {Metzner}\ \emph {et~al.}(2012)\citenamefont
  {Metzner}, \citenamefont {Salmhofer}, \citenamefont {Honerkamp},
  \citenamefont {Meden},\ and\ \citenamefont
  {Sch{\"{o}}nhammer}}]{Metzner2012a}%
  \BibitemOpen
  \bibfield  {author} {\bibinfo {author} {\bibfnamefont {W.}~\bibnamefont
  {Metzner}}, \bibinfo {author} {\bibfnamefont {M.}~\bibnamefont {Salmhofer}},
  \bibinfo {author} {\bibfnamefont {C.}~\bibnamefont {Honerkamp}}, \bibinfo
  {author} {\bibfnamefont {V.}~\bibnamefont {Meden}},\ and\ \bibinfo {author}
  {\bibfnamefont {K.}~\bibnamefont {Sch{\"{o}}nhammer}},\ }\bibfield  {title}
  {\bibinfo {title} {{Functional renormalization group approach to correlated
  fermion systems}},\ }\href {https://doi.org/10.1103/RevModPhys.84.299}
  {\bibfield  {journal} {\bibinfo  {journal} {Reviews of Modern Physics}\
  }\textbf {\bibinfo {volume} {84}},\ \bibinfo {pages} {299} (\bibinfo {year}
  {2012})}\BibitemShut {NoStop}%
\bibitem [{\citenamefont {Shtyk}\ \emph {et~al.}(2017)\citenamefont {Shtyk},
  \citenamefont {Goldstein},\ and\ \citenamefont
  {Chamon}}]{shtyk2017electrons}%
  \BibitemOpen
  \bibfield  {author} {\bibinfo {author} {\bibfnamefont {A.}~\bibnamefont
  {Shtyk}}, \bibinfo {author} {\bibfnamefont {G.}~\bibnamefont {Goldstein}},\
  and\ \bibinfo {author} {\bibfnamefont {C.}~\bibnamefont {Chamon}},\
  }\bibfield  {title} {\bibinfo {title} {Electrons at the monkey saddle: A
  multicritical lifshitz point},\ }\href
  {https://doi.org/10.1103/PhysRevB.95.035137} {\bibfield  {journal} {\bibinfo
  {journal} {Phys. Rev. B}\ }\textbf {\bibinfo {volume} {95}},\ \bibinfo
  {pages} {035137} (\bibinfo {year} {2017})}\BibitemShut {NoStop}%
\bibitem [{\citenamefont {Nandkishore}\ \emph {et~al.}(2012)\citenamefont
  {Nandkishore}, \citenamefont {Chern},\ and\ \citenamefont
  {Chubukov}}]{PhysRevLett.108.227204}%
  \BibitemOpen
  \bibfield  {author} {\bibinfo {author} {\bibfnamefont {R.}~\bibnamefont
  {Nandkishore}}, \bibinfo {author} {\bibfnamefont {G.-W.}\ \bibnamefont
  {Chern}},\ and\ \bibinfo {author} {\bibfnamefont {A.~V.}\ \bibnamefont
  {Chubukov}},\ }\bibfield  {title} {\bibinfo {title} {Itinerant half-metal
  spin-density-wave state on the hexagonal lattice},\ }\href
  {https://doi.org/10.1103/PhysRevLett.108.227204} {\bibfield  {journal}
  {\bibinfo  {journal} {Phys. Rev. Lett.}\ }\textbf {\bibinfo {volume} {108}},\
  \bibinfo {pages} {227204} (\bibinfo {year} {2012})}\BibitemShut {NoStop}%
\bibitem [{\citenamefont {Martin}\ and\ \citenamefont
  {Batista}(2008)}]{PhysRevLett.101.156402}%
  \BibitemOpen
  \bibfield  {author} {\bibinfo {author} {\bibfnamefont {I.}~\bibnamefont
  {Martin}}\ and\ \bibinfo {author} {\bibfnamefont {C.~D.}\ \bibnamefont
  {Batista}},\ }\bibfield  {title} {\bibinfo {title} {Itinerant electron-driven
  chiral magnetic ordering and spontaneous quantum hall effect in triangular
  lattice models},\ }\href {https://doi.org/10.1103/PhysRevLett.101.156402}
  {\bibfield  {journal} {\bibinfo  {journal} {Phys. Rev. Lett.}\ }\textbf
  {\bibinfo {volume} {101}},\ \bibinfo {pages} {156402} (\bibinfo {year}
  {2008})}\BibitemShut {NoStop}%
\bibitem [{\citenamefont {Honerkamp}(2003)}]{PhysRevB.68.104510}%
  \BibitemOpen
  \bibfield  {author} {\bibinfo {author} {\bibfnamefont {C.}~\bibnamefont
  {Honerkamp}},\ }\bibfield  {title} {\bibinfo {title} {Instabilities of
  interacting electrons on the triangular lattice},\ }\href
  {https://doi.org/10.1103/PhysRevB.68.104510} {\bibfield  {journal} {\bibinfo
  {journal} {Phys. Rev. B}\ }\textbf {\bibinfo {volume} {68}},\ \bibinfo
  {pages} {104510} (\bibinfo {year} {2003})}\BibitemShut {NoStop}%
\bibitem [{\citenamefont {Scherer}\ \emph {et~al.}(2021)\citenamefont
  {Scherer}, \citenamefont {Kennes},\ and\ \citenamefont
  {Classen}}]{scherer2021}%
  \BibitemOpen
  \bibfield  {author} {\bibinfo {author} {\bibfnamefont {M.~M.}\ \bibnamefont
  {Scherer}}, \bibinfo {author} {\bibfnamefont {D.~M.}\ \bibnamefont
  {Kennes}},\ and\ \bibinfo {author} {\bibfnamefont {L.}~\bibnamefont
  {Classen}},\ }\href@noop {} {\bibinfo {title} {$\mathcal{N}=4$ chiral
  superconductivity in moir\'e transition metal dichalcogenides}} (\bibinfo
  {year} {2021}),\ \Eprint {https://arxiv.org/abs/2108.11406} {arXiv:2108.11406
  [cond-mat.str-el]} \BibitemShut {NoStop}%
\bibitem [{\citenamefont {Gneist}\ \emph {et~al.}(2022)\citenamefont {Gneist},
  \citenamefont {Classen},\ and\ \citenamefont
  {Scherer}}]{gneist2022competing}%
  \BibitemOpen
  \bibfield  {author} {\bibinfo {author} {\bibfnamefont {N.}~\bibnamefont
  {Gneist}}, \bibinfo {author} {\bibfnamefont {L.}~\bibnamefont {Classen}},\
  and\ \bibinfo {author} {\bibfnamefont {M.~M.}\ \bibnamefont {Scherer}},\
  }\bibfield  {title} {\bibinfo {title} {Competing instabilities of the
  extended hubbard model on the triangular lattice: Truncated-unity functional
  renormalization group and application to moir\'e materials},\ }\href@noop {}
  {\bibfield  {journal} {\bibinfo  {journal} {arXiv preprint arXiv:2203.01226}\
  } (\bibinfo {year} {2022})}\BibitemShut {NoStop}%
\bibitem [{\citenamefont {Wietek}\ \emph {et~al.}(2021)\citenamefont {Wietek},
  \citenamefont {Rossi}, \citenamefont {\ifmmode~\check{S}\else
  \v{S}\fi{}imkovic}, \citenamefont {Klett}, \citenamefont {Hansmann},
  \citenamefont {Ferrero}, \citenamefont {Stoudenmire}, \citenamefont
  {Sch\"afer},\ and\ \citenamefont {Georges}}]{PhysRevX.11.041013}%
  \BibitemOpen
  \bibfield  {author} {\bibinfo {author} {\bibfnamefont {A.}~\bibnamefont
  {Wietek}}, \bibinfo {author} {\bibfnamefont {R.}~\bibnamefont {Rossi}},
  \bibinfo {author} {\bibfnamefont {F.}~\bibnamefont {\ifmmode~\check{S}\else
  \v{S}\fi{}imkovic}}, \bibinfo {author} {\bibfnamefont {M.}~\bibnamefont
  {Klett}}, \bibinfo {author} {\bibfnamefont {P.}~\bibnamefont {Hansmann}},
  \bibinfo {author} {\bibfnamefont {M.}~\bibnamefont {Ferrero}}, \bibinfo
  {author} {\bibfnamefont {E.~M.}\ \bibnamefont {Stoudenmire}}, \bibinfo
  {author} {\bibfnamefont {T.}~\bibnamefont {Sch\"afer}},\ and\ \bibinfo
  {author} {\bibfnamefont {A.}~\bibnamefont {Georges}},\ }\bibfield  {title}
  {\bibinfo {title} {Mott insulating states with competing orders in the
  triangular lattice hubbard model},\ }\href
  {https://doi.org/10.1103/PhysRevX.11.041013} {\bibfield  {journal} {\bibinfo
  {journal} {Phys. Rev. X}\ }\textbf {\bibinfo {volume} {11}},\ \bibinfo
  {pages} {041013} (\bibinfo {year} {2021})}\BibitemShut {NoStop}%
\bibitem [{\citenamefont {Sigrist}\ and\ \citenamefont
  {Ueda}(1991)}]{sigrist1991phenomenological}%
  \BibitemOpen
  \bibfield  {author} {\bibinfo {author} {\bibfnamefont {M.}~\bibnamefont
  {Sigrist}}\ and\ \bibinfo {author} {\bibfnamefont {K.}~\bibnamefont {Ueda}},\
  }\bibfield  {title} {\bibinfo {title} {Phenomenological theory of
  unconventional superconductivity},\ }\href
  {https://doi.org/10.1103/RevModPhys.63.239} {\bibfield  {journal} {\bibinfo
  {journal} {Rev. Mod. Phys.}\ }\textbf {\bibinfo {volume} {63}},\ \bibinfo
  {pages} {239} (\bibinfo {year} {1991})}\BibitemShut {NoStop}%
\bibitem [{Note1()}]{Note1}%
  \BibitemOpen
  \bibinfo {note} {Note that the symmetry is enhanced to $C_{6v}$ (equivalent
  to $D_{6h}$ due to the inherent two-dimensional nature of our model) for
  $\varphi =0$, where we find pairing instabilities in the $E_2$ ($d$-wave,
  $E_{2g}$), $B_1$ ($f$-wave, $B_{1u}$) and $A_2$ ($i$-wave, $A_{2g}$)
  representations.}\BibitemShut {Stop}%
\bibitem [{\citenamefont {Kapitulnik}(2015)}]{KAPITULNIK2015151}%
  \BibitemOpen
  \bibfield  {author} {\bibinfo {author} {\bibfnamefont {A.}~\bibnamefont
  {Kapitulnik}},\ }\bibfield  {title} {\bibinfo {title} {Notes on constraints
  for the observation of polar kerr effect in complex materials},\ }\href
  {https://doi.org/https://doi.org/10.1016/j.physb.2014.11.059} {\bibfield
  {journal} {\bibinfo  {journal} {Physica B: Condensed Matter}\ }\textbf
  {\bibinfo {volume} {460}},\ \bibinfo {pages} {151} (\bibinfo {year}
  {2015})},\ \bibinfo {note} {special Issue on Electronic Crystals
  (ECRYS-2014)}\BibitemShut {NoStop}%
\bibitem [{\citenamefont {Mahyari}\ \emph {et~al.}(2014)\citenamefont
  {Mahyari}, \citenamefont {Cannell}, \citenamefont {Gomez}, \citenamefont
  {Tezok}, \citenamefont {Zelati}, \citenamefont {de~Mello}, \citenamefont
  {Yan}, \citenamefont {Mandrus},\ and\ \citenamefont
  {Sonier}}]{PhysRevB.89.020502}%
  \BibitemOpen
  \bibfield  {author} {\bibinfo {author} {\bibfnamefont {Z.~L.}\ \bibnamefont
  {Mahyari}}, \bibinfo {author} {\bibfnamefont {A.}~\bibnamefont {Cannell}},
  \bibinfo {author} {\bibfnamefont {C.}~\bibnamefont {Gomez}}, \bibinfo
  {author} {\bibfnamefont {S.}~\bibnamefont {Tezok}}, \bibinfo {author}
  {\bibfnamefont {A.}~\bibnamefont {Zelati}}, \bibinfo {author} {\bibfnamefont
  {E.~V.~L.}\ \bibnamefont {de~Mello}}, \bibinfo {author} {\bibfnamefont
  {J.-Q.}\ \bibnamefont {Yan}}, \bibinfo {author} {\bibfnamefont {D.~G.}\
  \bibnamefont {Mandrus}},\ and\ \bibinfo {author} {\bibfnamefont {J.~E.}\
  \bibnamefont {Sonier}},\ }\bibfield  {title} {\bibinfo {title} {Zero-field
  $\ensuremath{\mu}$sr search for a time-reversal-symmetry-breaking mixed
  pairing state in superconducting
  ba${}_{1\ensuremath{-}x}$k${}_{x}$fe${}_{2}$as${}_{2}$},\ }\href
  {https://doi.org/10.1103/PhysRevB.89.020502} {\bibfield  {journal} {\bibinfo
  {journal} {Phys. Rev. B}\ }\textbf {\bibinfo {volume} {89}},\ \bibinfo
  {pages} {020502} (\bibinfo {year} {2014})}\BibitemShut {NoStop}%
\bibitem [{\citenamefont {Kozii}\ \emph {et~al.}(2019)\citenamefont {Kozii},
  \citenamefont {Isobe}, \citenamefont {Venderbos},\ and\ \citenamefont
  {Fu}}]{PhysRevB.99.144507}%
  \BibitemOpen
  \bibfield  {author} {\bibinfo {author} {\bibfnamefont {V.}~\bibnamefont
  {Kozii}}, \bibinfo {author} {\bibfnamefont {H.}~\bibnamefont {Isobe}},
  \bibinfo {author} {\bibfnamefont {J.~W.~F.}\ \bibnamefont {Venderbos}},\ and\
  \bibinfo {author} {\bibfnamefont {L.}~\bibnamefont {Fu}},\ }\bibfield
  {title} {\bibinfo {title} {Nematic superconductivity stabilized by density
  wave fluctuations: Possible application to twisted bilayer graphene},\ }\href
  {https://doi.org/10.1103/PhysRevB.99.144507} {\bibfield  {journal} {\bibinfo
  {journal} {Phys. Rev. B}\ }\textbf {\bibinfo {volume} {99}},\ \bibinfo
  {pages} {144507} (\bibinfo {year} {2019})}\BibitemShut {NoStop}%
\bibitem [{\citenamefont {Chichinadze}\ \emph {et~al.}(2020)\citenamefont
  {Chichinadze}, \citenamefont {Classen},\ and\ \citenamefont
  {Chubukov}}]{PhysRevB.101.224513}%
  \BibitemOpen
  \bibfield  {author} {\bibinfo {author} {\bibfnamefont {D.~V.}\ \bibnamefont
  {Chichinadze}}, \bibinfo {author} {\bibfnamefont {L.}~\bibnamefont
  {Classen}},\ and\ \bibinfo {author} {\bibfnamefont {A.~V.}\ \bibnamefont
  {Chubukov}},\ }\bibfield  {title} {\bibinfo {title} {Nematic
  superconductivity in twisted bilayer graphene},\ }\href
  {https://doi.org/10.1103/PhysRevB.101.224513} {\bibfield  {journal} {\bibinfo
   {journal} {Phys. Rev. B}\ }\textbf {\bibinfo {volume} {101}},\ \bibinfo
  {pages} {224513} (\bibinfo {year} {2020})}\BibitemShut {NoStop}%
\bibitem [{\citenamefont {Yonezawa}\ \emph {et~al.}(2017)\citenamefont
  {Yonezawa}, \citenamefont {Tajiri}, \citenamefont {Nakata}, \citenamefont
  {Nagai}, \citenamefont {Wang}, \citenamefont {Segawa}, \citenamefont {Ando},\
  and\ \citenamefont {Maeno}}]{Yonezawa2017}%
  \BibitemOpen
  \bibfield  {author} {\bibinfo {author} {\bibfnamefont {S.}~\bibnamefont
  {Yonezawa}}, \bibinfo {author} {\bibfnamefont {K.}~\bibnamefont {Tajiri}},
  \bibinfo {author} {\bibfnamefont {S.}~\bibnamefont {Nakata}}, \bibinfo
  {author} {\bibfnamefont {Y.}~\bibnamefont {Nagai}}, \bibinfo {author}
  {\bibfnamefont {Z.}~\bibnamefont {Wang}}, \bibinfo {author} {\bibfnamefont
  {K.}~\bibnamefont {Segawa}}, \bibinfo {author} {\bibfnamefont
  {Y.}~\bibnamefont {Ando}},\ and\ \bibinfo {author} {\bibfnamefont
  {Y.}~\bibnamefont {Maeno}},\ }\bibfield  {title} {\bibinfo {title}
  {Thermodynamic evidence for nematic superconductivity in cuxbi2se3},\ }\href
  {https://doi.org/10.1038/nphys3907} {\bibfield  {journal} {\bibinfo
  {journal} {Nature Physics}\ }\textbf {\bibinfo {volume} {13}},\ \bibinfo
  {pages} {123} (\bibinfo {year} {2017})}\BibitemShut {NoStop}%
\bibitem [{\citenamefont {Shen}\ \emph {et~al.}(2017)\citenamefont {Shen},
  \citenamefont {He}, \citenamefont {Yuan}, \citenamefont {Huang},
  \citenamefont {Cho}, \citenamefont {Lee}, \citenamefont {Hor}, \citenamefont
  {Law},\ and\ \citenamefont {Lortz}}]{Shen2017}%
  \BibitemOpen
  \bibfield  {author} {\bibinfo {author} {\bibfnamefont {J.}~\bibnamefont
  {Shen}}, \bibinfo {author} {\bibfnamefont {W.-Y.}\ \bibnamefont {He}},
  \bibinfo {author} {\bibfnamefont {N.~F.~Q.}\ \bibnamefont {Yuan}}, \bibinfo
  {author} {\bibfnamefont {Z.}~\bibnamefont {Huang}}, \bibinfo {author}
  {\bibfnamefont {C.-w.}\ \bibnamefont {Cho}}, \bibinfo {author} {\bibfnamefont
  {S.~H.}\ \bibnamefont {Lee}}, \bibinfo {author} {\bibfnamefont {Y.~S.}\
  \bibnamefont {Hor}}, \bibinfo {author} {\bibfnamefont {K.~T.}\ \bibnamefont
  {Law}},\ and\ \bibinfo {author} {\bibfnamefont {R.}~\bibnamefont {Lortz}},\
  }\bibfield  {title} {\bibinfo {title} {Nematic topological superconducting
  phase in nb-doped bi2se3},\ }\href
  {https://doi.org/10.1038/s41535-017-0064-1} {\bibfield  {journal} {\bibinfo
  {journal} {npj Quantum Materials}\ }\textbf {\bibinfo {volume} {2}},\
  \bibinfo {pages} {59} (\bibinfo {year} {2017})}\BibitemShut {NoStop}%
\bibitem [{\citenamefont {Cao}\ \emph {et~al.}(2021{\natexlab{b}})\citenamefont
  {Cao}, \citenamefont {Rodan-Legrain}, \citenamefont {Park}, \citenamefont
  {Yuan}, \citenamefont {Watanabe}, \citenamefont {Taniguchi}, \citenamefont
  {Fernandes}, \citenamefont {Fu},\ and\ \citenamefont
  {Jarillo-Herrero}}]{doi:10.1126/science.abc2836}%
  \BibitemOpen
  \bibfield  {author} {\bibinfo {author} {\bibfnamefont {Y.}~\bibnamefont
  {Cao}}, \bibinfo {author} {\bibfnamefont {D.}~\bibnamefont {Rodan-Legrain}},
  \bibinfo {author} {\bibfnamefont {J.~M.}\ \bibnamefont {Park}}, \bibinfo
  {author} {\bibfnamefont {N.~F.~Q.}\ \bibnamefont {Yuan}}, \bibinfo {author}
  {\bibfnamefont {K.}~\bibnamefont {Watanabe}}, \bibinfo {author}
  {\bibfnamefont {T.}~\bibnamefont {Taniguchi}}, \bibinfo {author}
  {\bibfnamefont {R.~M.}\ \bibnamefont {Fernandes}}, \bibinfo {author}
  {\bibfnamefont {L.}~\bibnamefont {Fu}},\ and\ \bibinfo {author}
  {\bibfnamefont {P.}~\bibnamefont {Jarillo-Herrero}},\ }\bibfield  {title}
  {\bibinfo {title} {Nematicity and competing orders in superconducting
  magic-angle graphene},\ }\href {https://doi.org/10.1126/science.abc2836}
  {\bibfield  {journal} {\bibinfo  {journal} {Science}\ }\textbf {\bibinfo
  {volume} {372}},\ \bibinfo {pages} {264} (\bibinfo {year}
  {2021}{\natexlab{b}})},\ \Eprint
  {https://arxiv.org/abs/https://www.science.org/doi/pdf/10.1126/science.abc2836}
  {https://www.science.org/doi/pdf/10.1126/science.abc2836} \BibitemShut
  {NoStop}%
\bibitem [{\citenamefont {Asaba}\ \emph {et~al.}(2017)\citenamefont {Asaba},
  \citenamefont {Lawson}, \citenamefont {Tinsman}, \citenamefont {Chen},
  \citenamefont {Corbae}, \citenamefont {Li}, \citenamefont {Qiu},
  \citenamefont {Hor}, \citenamefont {Fu},\ and\ \citenamefont
  {Li}}]{PhysRevX.7.011009}%
  \BibitemOpen
  \bibfield  {author} {\bibinfo {author} {\bibfnamefont {T.}~\bibnamefont
  {Asaba}}, \bibinfo {author} {\bibfnamefont {B.~J.}\ \bibnamefont {Lawson}},
  \bibinfo {author} {\bibfnamefont {C.}~\bibnamefont {Tinsman}}, \bibinfo
  {author} {\bibfnamefont {L.}~\bibnamefont {Chen}}, \bibinfo {author}
  {\bibfnamefont {P.}~\bibnamefont {Corbae}}, \bibinfo {author} {\bibfnamefont
  {G.}~\bibnamefont {Li}}, \bibinfo {author} {\bibfnamefont {Y.}~\bibnamefont
  {Qiu}}, \bibinfo {author} {\bibfnamefont {Y.~S.}\ \bibnamefont {Hor}},
  \bibinfo {author} {\bibfnamefont {L.}~\bibnamefont {Fu}},\ and\ \bibinfo
  {author} {\bibfnamefont {L.}~\bibnamefont {Li}},\ }\bibfield  {title}
  {\bibinfo {title} {Rotational symmetry breaking in a trigonal superconductor
  nb-doped ${\mathrm{bi}}_{2}{\mathrm{se}}_{3}$},\ }\href
  {https://doi.org/10.1103/PhysRevX.7.011009} {\bibfield  {journal} {\bibinfo
  {journal} {Phys. Rev. X}\ }\textbf {\bibinfo {volume} {7}},\ \bibinfo {pages}
  {011009} (\bibinfo {year} {2017})}\BibitemShut {NoStop}%
\bibitem [{\citenamefont {Klebl}\ \emph
  {et~al.}(2022{\natexlab{b}})\citenamefont {Klebl}, \citenamefont {Xu},
  \citenamefont {Fischer}, \citenamefont {Xian}, \citenamefont {Claassen},
  \citenamefont {Rubio},\ and\ \citenamefont {Kennes}}]{klebl2022moire}%
  \BibitemOpen
  \bibfield  {author} {\bibinfo {author} {\bibfnamefont {L.}~\bibnamefont
  {Klebl}}, \bibinfo {author} {\bibfnamefont {Q.}~\bibnamefont {Xu}}, \bibinfo
  {author} {\bibfnamefont {A.}~\bibnamefont {Fischer}}, \bibinfo {author}
  {\bibfnamefont {L.}~\bibnamefont {Xian}}, \bibinfo {author} {\bibfnamefont
  {M.}~\bibnamefont {Claassen}}, \bibinfo {author} {\bibfnamefont
  {A.}~\bibnamefont {Rubio}},\ and\ \bibinfo {author} {\bibfnamefont {D.~M.}\
  \bibnamefont {Kennes}},\ }\bibfield  {title} {\bibinfo {title} {Moir{\'{e}}
  engineering of spin{\textendash}orbit coupling in twisted platinum
  diselenide},\ }\href {https://doi.org/10.1088/2516-1075/ac49f5} {\bibfield
  {journal} {\bibinfo  {journal} {Electronic Structure}\ }\textbf {\bibinfo
  {volume} {4}},\ \bibinfo {pages} {014004} (\bibinfo {year}
  {2022}{\natexlab{b}})}\BibitemShut {NoStop}%
\end{thebibliography}%

\arxivSubmit{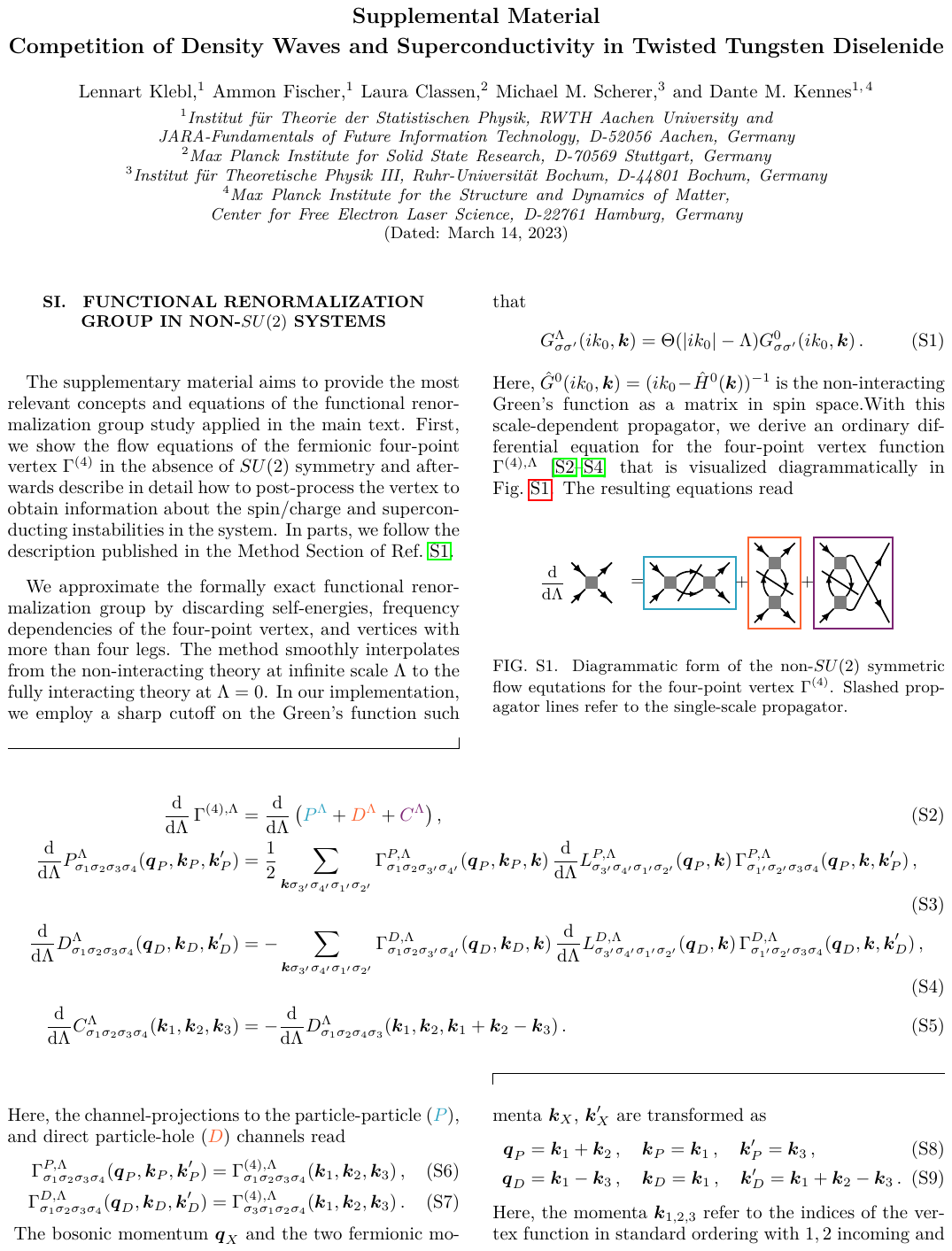}{6}

\end{document}